\documentclass[11pt,a4paper]{article}
\pdfoutput=1
\usepackage{graphicx}

\usepackage{jheppub}
\usepackage{amsmath}
\usepackage{mathtools}

\newcounter{fig}

\newcommand{\nn}{\nonumber}

\newcommand{\rs}{r_\star}
\newcommand{\rst}{r_{\star t}}
\newcommand{\rsc}{r_{\star c}}
\newcommand{\rsb}{r_{\star b}}
\newcommand{\F}{{}_2F_1}

\newcommand{\mfu}{\mathfrak{u}}
\newcommand{\mfw}{\mathfrak{w}}
\newcommand{\mfq}{\mathfrak{q}}

\newcommand\ft[2]{{\textstyle\frac{#1}{#2}}}

\def\[{\left [}
\def\]{\right ]}
\def\({\left (}
\def\){\right )}

\def\r{\rho}

\def\r2{\sqrt{2}}

\newcommand{\bbibitem}[1]{\bibitem{#1}\marginpar{#1}}

\def\Label#1{\label{#1}
  \smash{\hbox to0pt{\raise1ex\hbox{\tiny[#1]}\hss}}}
\def\noLabels{\let\Label=\label}
\def\nobbibitem{\let\bbibitem=\bibitem}

\newcommand{\bea}{\begin{eqnarray}}
\newcommand{\eea}{\end{eqnarray}}

\newcommand{\beq} {\begin{equation}}
\newcommand{\eeq} {\end{equation}}
\newcommand{\beqa} {\begin{eqnarray}}
\newcommand{\eeqa} {\end{eqnarray}}

\newcommand{\beqn}{\begin{eqnarray}}
\newcommand{\eeqn}{\end{eqnarray}}

\renewcommand{\thepage}{\arabic{page}}
\title{Correlation functions in theories\\with Lifshitz scaling}
                                           \author[a,b]{Ville Ker\"{a}nen,}
                                           \author[c]{Watse Sybesma,}
					 \author[c]{Phillip Szepietowski}
                                           \author[d,e]{and Larus Thorlacius}
                                           \affiliation[a]{\it Helsinki Institute of Physics,\\
                                          \it P.O.Box 64, FI-00014 University of Helsinki, Finland}
                                           \affiliation[b]{\it Department of Physics,\\
                                           \it P.O.Box 64, FIN-00014 University of Helsinki, Finland}
                                           \affiliation[c]{\it Institute for Theoretical Physics and Center for Extreme Matter and Emergent Phenomena,\\
                                           \it Utrecht University, Princetonplein 5, 3584 CC Utrecht, the Netherlands}
					 \affiliation[d]{\it University of Iceland, Science Institute,\\
                                           \it Dunhaga 3, IS-107 Reykjavik, Iceland}
					 \affiliation[e]{The Oskar Klein Centre for Cosmoparticle Physics\\
					 \it Department of Physics, Stockholm University,\\
                                           \it AlbaNova University Centre, SE-106 91 Stockholm, Sweden}
                                           \emailAdd{vkeranen1@gmail.com}
                                           \emailAdd{z.w.sybesma@uu.nl}
                                           \emailAdd{p.g.szepietowski@uu.nl}
					 \emailAdd{lth@hi.is}
                                           \abstract{The 2+1 dimensional quantum Lifshitz model can be generalised 
to a class of higher dimensional free field theories that exhibit
Lifshitz scaling. When the dynamical critical exponent equals the
number of spatial dimensions, equal time correlation functions
of scaling operators in the generalised quantum Lifshitz model 
are given by a d-dimensional higher-derivative conformal field theory. 
Autocorrelation functions in the generalised quantum Lifshitz model in 
any number of dimensions can on the other hand be expressed in terms 
of autocorrelation functions of a two-dimensional conformal field theory. 
This also holds for autocorrelation functions in a strongly coupled 
Lifshitz field theory with a holographic dual of Einstein-Maxwell-dilaton 
type. The map to a two-dimensional conformal field theory extends to 
autocorrelation functions in thermal states and out-of-equilbrium states 
preserving symmetry under spatial translations and rotations in both 
types of Lifshitz models. Furthermore, the spectrum of quasinormal 
modes of scalar field perturbations in Lifshitz black hole backgrounds 
can be obtained analytically at low spatial momenta and exhibits a 
linear dispersion relation at z = d.  At high momentum, the mode spectrum 
can be obtained in a WKB approximation and displays very different 
behaviour compared to holographic duals of conformal field theories. 
This has implications for thermalisation in strongly coupled Lifshitz field theories with $z>1$.}

                                           \arxivnumber{1611.09371}
                                           
\begin{document}
                                           \maketitle

\section{Introduction}

A second order quantum phase transition is characterized by a dynamical critical scaling 
exponent $z$, which dictates the form of the scaling symmetry present at the zero temperature
critical point,
\beq
\vec{x}\rightarrow \lambda \vec{x},\quad \tau\rightarrow \lambda^z \tau\,,
\label{xtscaling}
\eeq
where $\vec{x}=(x^1,\ldots,x^d)$ are spatial coordinates and $\tau$ is 
Euclidean time. In relativistic systems $z=1$ and the scaling symmetry is extended 
to the full conformal symmetry. Generic quantum critical points in non-relativistic systems,
on the other hand, do not have conformal symmetry and exhibit anisotropic scaling between 
time and the spatial directions with $z>1$, commonly referred to as Lifshitz scaling. 
In this paper we study correlation functions of scaling operators in theories with Lifshitz 
scaling in $d\geq 2$ spatial dimensions. We pay particular attention
to so-called autocorrelators, {\it i.e.} correlation functions where all the operators are
inserted at the same spatial point but at different times,
\beq
\langle\mathcal{O}_1(\vec{x},t_1)\mathcal{O}_2(\vec{x},t_2)...
\mathcal{O}_n(\vec{x},t_n)\rangle\,.
\eeq
Autocorrelators carry information 
about the energy spectrum of the model in question and provide a useful diagnostic of
time evolution in out of equilibrium configurations.

We begin in Section~\ref{quantumlifshitz} by introducing a set of free field theories in
$d{+}1$ dimensions that realise Lifshitz scaling with arbitrary dynamical critical exponent. 
These theories include the well known $2{+}1$-dimensional 
quantum Lifshitz model as a special case with $z=d=2$ and provide a particularly 
simple setting to study quantum critical behaviour at generic integer values of $z$. 
Many interesting properties of the original quantum Lifshitz model survive in the more 
general free field theories when the dynamical critical exponent $z$ equals the number 
of spatial dimensions and we refer to the models with $z=d$ as 
 {\it generalised quantum Lifshitz models}. For instance, it is well known that the 
ground state correlation functions of scaling operators at equal time in the original 
quantum Lifshitz model can be expressed in terms of correlation functions of a 
two-dimensional Euclidean field theory with conformal symmetry~\cite{Ardonne:2003wa}. 
We briefly review this construction and then show how the connection to conformal field 
theory extends to $d{+}1$-dimensions, provided that $z=d$, in which case the 
connection is to a higher-derivative free field CFT in $d$ dimensions. 

There is a further connection to conformal field theory, which is only realised in the
time domain. It turns out that autocorrelators in the generalised quantum Lifshitz model 
in any number of dimensions can be expressed in terms of autocorrelators of a 
two-dimensional conformal field theory when $z=d$. This holds for autocorrelators evaluated in 
any Gaussian state that is symmetric under spatial translations and rotations, and as a non-trivial
example we work out autocorrelators of monopole operators in a thermal state.

Since the generalised quantum Lifshitz model is a free field theory, it is perhaps not that 
surprising that its correlation functions take a simple form. It turns out, however, that the 
connection to two-dimensional conformal field theory persists when one considers 
autocorrelators in a strongly coupled field theory with a holographic dual that exhibits 
Lifshitz scaling. In Section~\ref{holomodel} we introduce a holographic model that realises 
Lifshitz scaling with generic $z> 1$ and evaluate vacuum two-point functions of operators 
with large scaling dimension. Analytic expressions are easily obtained for the equal time 
two-point function and the two-point autocorrelator at general $d$ and $z$. For the special case 
of $z=2$ we find a parametrized solution for the full vacuum two-point function at 
arbitrary spatial and temporal separation on the boundary and compare it to the 
corresponding correlation function in the quantum Lifshitz model. 
We then turn our attention to more general autocorrelators in the holographic model 
and uncover a quantitative relation to autocorrelators in a two-dimensional conformal 
field theory. In particular, we show how the form of three-point autocorrelators 
at general $d$ and $z$ is constrained by an underlying two-dimensional extended symmetry. 

In Section~\ref{thermalcorrelators} we restrict to integer values of $z=d$ and show how 
autocorrelators in a thermal state in the holographic model can be expressed in 
terms of thermal autocorrelators of a two-dimensional conformal field theory. 
The resulting thermal two- and three-point autocorrelators in the holographic theory 
thus have the same scaling form as those of the generalised quantum Lifshitz model. 

In Section~\ref{sec:finite-Delta} we venture outside the time domain and consider 
quasinormal modes of massive scalar field fluctuations at finite momentum. 
In momentum space we can go beyond the geodesic approximation and find that the special behaviour at $z=d$ persists at small momentum and any operator scaling dimension. 
In Appendix~\ref{app:QNM} we expand on the numerical and analytic methods we use to 
study the quasinormal mode spectrum and extend our discussion to include quasinormal 
modes at any $z$, $d$ and high momentum.

The connection between generalised Lifshitz models at $z=d$ and two-dimensional 
conformal field theory extends to more general states that are invariant under spatial translations
and rotations. This includes time-dependent states dual to Lifshitz-Vaidya spacetimes
in the holographic models and quench states in the generalised quantum Lifshitz model, 
as briefly illustrated in Section~\ref{nonequilibrium}.

We conclude our discussion in Section~\ref{discussion} and some technical 
results that are referred to in the main text are worked out in the appendices.

\section{The generalised quantum Lifshitz model}
\label{quantumlifshitz}
Our starting point is a class of $d{+}1$-dimensional quantum field theories exhibiting 
Lifshitz scaling with integer valued dynamical critical exponent $z$. 
They are governed by the following (Euclidean) action
\beq
S=\frac{1}{2}\int d^dxd\tau\Big[(\partial_{\tau}\chi)^2
+\kappa^2(\nabla^z\chi)^2\Big],
\label{eq:quantumlifshitz}
\eeq
where $\kappa>0$ is a constant and we are using a shorthand notation,
\beq
\nabla^z\chi = \left\{
\begin{array}{cll}
(\nabla^2)^k\chi & \textrm{if} & z=2k \,, \\
(\nabla^2)^k\vec\nabla\chi & \textrm{if} & z=2k{+}1 \,.
\end{array}
\right.
\eeq
These models generalise the well known quantum Lifshitz model~\cite{Ardonne:2003wa}, 
which has $z=2$ in $2{+}1$ dimensions, and allow us to systematically explore the 
behaviour of various physical quantities at different values of the dynamical critical
exponent in a controlled free field theory setting. When there are explicit analytic results  
available one can even consider formally extending $z$ to non-integer values.
In the following we will mainly be concerned with theories where $z=d$, which
includes the quantum Lifshitz model as a special case. The connection 
to conformal field theory found in the quantum Lifshitz model is retained at general
$z=d>2$ and accordingly we refer to these theories as {\it generalised quantum 
Lifshitz models}.

The generalised monopole operators (or just monopole operators for short),
\beq
\mathcal{O}_{\alpha}(\vec{x},t)=e^{i\alpha\chi(\vec{x},t)},
\label{monopoleops}
\eeq
with $\alpha\in \mathbf{R}$, turn out to have simple scaling properties in the 
generalised quantum Lifshitz model and
we will focus our attention on their correlation functions.
The name monopole operator arises from a dual gauge field representation of the 
original quantum Lifshitz model, where such operators with $\alpha=\pm 2\pi$ create 
magnetic monopoles in three-dimensional Euclidean space~\cite{Ardonne:2003wa}. 

\subsection{Equal time correlation functions}\label{equaltimecorrelators}
The ground state wave functional of the quantum Lifshitz model with $z=d=2$ is given 
by the exponential of a Euclidean action of a free $1{+}1$-dimensional 
conformal field theory (CFT) \cite{Ardonne:2003wa}. In Appendix~\ref{appendixA} we 
outline how this statement can be generalised to arbitrary $z=d$ by expressing the ground 
state wave functional in terms of a free $d$-dimensional (higher-derivative) CFT
of a type studied recently in~\cite{Brust:2016gjy}. The
connection to a CFT can also be obtained by showing that equal time correlation 
functions of monopole operators in the generalised quantum Lifshitz model  
have the form of CFT correlation functions,
\beq
\langle \mathcal{O}_{\alpha_1}(\vec{x}_1,\tau)\ldots
\mathcal{O}_{\alpha_n}(\vec{x}_n,\tau)\rangle
=\big\langle e^{i\alpha_1\chi(\vec{x}_1)}\ldots e^{i\alpha_n\chi(\vec{x}_n)}\big\rangle_{CFT}.
\label{eq:lifshitzcft}
\eeq
Consider the vacuum two-point function of the $\chi$ field,
\beq
G_E(\vec{x}_1,\tau_1;\vec{x}_2,\tau_2)=\int \frac{d\omega d^dp}{(2\pi)^{d+1}}
\frac{e^{-i\omega(\tau_1-\tau_2)-i\vec{p}\cdot (\vec{x}_1-\vec{x}_2)}}{\omega^2+\kappa^2 p^{2z}}\,,
\label{eq:twopointfn}
\eeq
where $p=|\vec p|$ and we let $z$ take a general integer value for now. 
The integral over $\omega$ is easily performed by closing the integration contour around 
the upper or lower half of the complex $\omega$-plane, depending on the sign of 
$\tau_{12}\equiv\tau_1-\tau_2$. Either way, a single residue at 
$\omega=\pm i \kappa p^z$ is picked out, giving
\beq
G_E(\vec{x}_1,\tau_1;\vec{x}_2,\tau_2)=
\frac{1}{2\kappa}\int\frac{d^dp}{(2\pi)^d}
\frac{1}{p^z}e^{-\kappa |\tau_{12}|p^z-i\vec{p}\cdot \vec{x}_{12}} ,
\label{full2ptfcn}
\eeq
where $\vec{x}_{12}\equiv \vec{x}_1-\vec{x}_2$. 
The remaining momentum integral is convergent for $z<d$, logarithmically divergent 
for $z=d$, and power-law divergent for $z>d$. 

For equal time correlation functions we set $\tau_1=\tau_2$ 
in \eqref{full2ptfcn} and then the two-point function 
becomes formally identical to the two-point function 
of a free scalar field in $d$-dimensions with the action
\beq
S= \kappa\int d^dx\phi(-\nabla^2)^{z/2}\phi\,.
\label{equaltimeaction}
\eeq
At $z=d$ with $d$ an even integer, this is one of the higher-derivative free
scalar CFTs considered in \cite{Brust:2016gjy}.\footnote{In odd numbered spatial 
dimensions the action \eqref{equaltimeaction} appears non-analytic in momentum space
but we can still obtain a free CFT at $z=d$ at odd $d$. A free field theory is fully determined 
by its two-point functions and \eqref{full2ptfcn} with $\tau_1=\tau_2$ supplies a well-defined 
two-point function for the $\phi$ field for any integer $d$.}

Equal time correlation functions of monopole operators are obtained by applying 
Wick's theorem, and the connection (\ref{eq:lifshitzcft}) follows from the equality 
of the two-point functions of the elementary fields at equal time in the generalised 
quantum Lifshitz model and the free $d$-dimensional CFT.
So far, we have written down formal expressions which need to be supplemented by
prescriptions for both IR and UV regularisation. To address such issues, and 
in order to establish some notation for later use, it is useful to work out some explicit
examples, starting with the equal time correlation function of two monopole operators, 
\begin{eqnarray}
\langle \mathcal{O}_{\alpha_1}(\vec{x}_1,\tau)
\mathcal{O}_{\alpha_2}(\vec{x}_2,\tau)\rangle
&=&\Big\langle e^{i\alpha_1\chi(\vec{x}_1)}e^{i\alpha_2\chi(\vec{x}_2)}\Big\rangle
\nonumber \\
&=&e^{-\frac12 \alpha_1^2 G_E(\vec{x}_1;\vec{x}_1)
-\frac12 \alpha_2^2 G_E(\vec{x}_2;\vec{x}_2)
-\alpha_1\alpha_2 G_E(\vec{x}_1;\vec{x}_2)} .
\label{eq:wickresult}
\end{eqnarray}
The equal time $G_E(\vec{x}_1;\vec{x}_2)$ is given by (\ref{full2ptfcn}) with $z=d$ 
and $\tau_1=\tau_2$, 
\begin{eqnarray}
G_E(\vec{x}_1;\vec{x}_2)
&=&\frac{1}{2\kappa}\int\frac{d^dp}{(2\pi)^d}
\frac{e^{-i\vec{p}\cdot\vec{x}_{12}}}{p^d} \nonumber \\
&=&
\frac{\text{Vol}(S^{d-2})}{2\kappa(2\pi)^{d}}
\int_0^\infty dp \int_0^\pi d\theta  (\sin\theta)^{d-2} 
\frac{e^{-i p |\vec{x}_{12}|  \cos\theta}}{p+\mu} \,,
\end{eqnarray}
where the parameter $\mu>0$, introduced to 
regulate the infrared divergence in the integral over $p$, is to be sent to
zero at the end of the calculation.

The integral over $\theta$ evaluates to a Bessel function,
\begin{equation}
\int_0^\pi d\theta (\sin \theta)^{d-2} e^{-i p |\vec{x}_{12}|  \cos \theta} 
= 2^{(d{-}2)/2}\sqrt{\pi}\Gamma(\ft{d-1}{2})\big(p |\vec{x}_{12}|\big)^{1-\ft{d}{2}} 
J_{\ft{d}{2}-1}(p |\vec{x}_{12}|), \nonumber
\end{equation}
and using $\text{Vol}(S^{d-2}) =2\pi^{(d-1)/2}/\Gamma(\frac{d-1}{2})$ we obtain
\begin{eqnarray}
G_E(\vec{x}_1;\vec{x}_2)&=&\frac{1}{2\kappa(2\pi)^{d/2}}
\int_0^\infty dy\, \frac{y^{1-\ft{d}{2}}}{y+\mu |\vec{x}_{12}|}\,J_{\ft{d}{2}-1}(y) \nonumber\\
&=&-\,\frac{1}{2^d\pi^{d/2}\Gamma(d/2)\kappa}\big(\log(\mu|\vec{x}_{12}|)+c_d+\ldots\big),
\end{eqnarray}
where $c_d$ is a finite constant and $\ldots$ denotes terms that vanish in the 
limit $\mu\rightarrow 0$. 
Inserting this expression for $G_E(\vec{x}_1;\vec{x}_2)$ into \eqref{eq:wickresult} then gives
\beq
\big\langle \mathcal{O}_{\alpha_1}(\vec{x}_1,\tau)
\mathcal{O}_{\alpha_2}(\vec{x}_2,\tau)\big\rangle
=\big(\mu e^{c_d}\big)^{\frac{(\alpha_1+\alpha_2)^2}{2^{d+1}\pi^{d/2}\Gamma(d/2) \kappa}}\,
\epsilon^{\frac{\alpha^2_1+\alpha^2_2}{2^{d+1}\pi^{d/2}\Gamma(d/2) \kappa}}\,
|\vec{x}_{12}|^{\frac{\alpha_1\alpha_2}{2^{d}\pi^{d/2}\Gamma(d/2) \kappa}} \,,
\eeq
where we have introduced an ultraviolet cutoff,
\beq
\vert \vec{x}_{ij}\vert\geq\epsilon\,,
\eeq
to regulate the self-contractions in Wick's theorem.
As a result, the equal time two-point function of monopole operators is independent of the 
infrared regulator when $\alpha_1+\alpha_2=0$ but vanishes when $\mu\rightarrow 0$ 
unless this condition on the charges is satisfied. The dependence on the 
ultraviolet cutoff can be absorbed into a renormalisation of the monopole operator when 
viewed as a composite operator, 
\beq
\mathcal{O}_{\alpha}^R(\vec{x},\tau)\equiv 
\epsilon^{-\Delta}e^{i\alpha\chi(\vec{x}_i,\tau)} \,,
\label{eq:renormoperator}
\eeq
of scaling dimension 
\begin{equation}\label{eq:scalingdimension}
	\Delta=\frac{\alpha^2}{2^{d+1}\pi^{d/2} \Gamma\left(\frac{d}{2}\right)\kappa}\,.
\end{equation}
The equal time two-point function of renormalized monopole operators then reduces 
to the usual scaling form,
\beq
\langle \mathcal{O}_{\alpha}^R(\vec{x}_1,\tau)
\mathcal{O}_{-\alpha}^R(\vec{x}_2,\tau)\rangle
=|\vec{x}_{12}|^{-2\Delta}.
\label{eq:vacuum2pt}
\eeq

The equal time correlation function of three monopole operators is 
obtained in a similar fashion,
\begin{align}
\langle& \mathcal{O}_{\alpha_1}(\vec{x}_1,\tau)
\mathcal{O}_{\alpha_2}(\vec{x}_2,\tau)
\mathcal{O}_{\alpha_3}(\vec{x}_3,\tau)\rangle \nonumber\\
&= e^{-\frac12 \sum_{i,j=1}^3\alpha_i\alpha_j G_E(\vec{x}_i;\vec{x}_j)} \\
&=\big(\mu e^{c_d}\big)^{\frac{(\alpha_1+\alpha_2+\alpha_3)^2}
{2^{d+1}\pi^{d/2}\Gamma(d/2) \kappa}}\epsilon^{\Delta_1+\Delta_2+\Delta_3}
|\vec{x}_{12}|^{\Delta_3-\Delta_1-\Delta_2}
|\vec{x}_{13}|^{\Delta_2-\Delta_3-\Delta_1}
|\vec{x}_{23}|^{\Delta_1-\Delta_2-\Delta_3} \,.
\nonumber
\label{eq:equaltime3pt}
\end{align}
This expression vanishes in the $\mu\rightarrow 0$ limit unless 
$\alpha_1+\alpha_2+\alpha_3=0$ but when the sum of charges
is zero the equal time three-point function of renormalised monopole
operators reduces to a standard CFT form,
\beq
\langle \mathcal{O}^R_{\alpha_1}(\vec{x}_1,\tau)
\mathcal{O}^R_{\alpha_2}(\vec{x}_2,\tau)
\mathcal{O}^R_{\alpha_3}(\vec{x}_3,\tau)\rangle
=|\vec{x}_{12}|^{\Delta_3-\Delta_1-\Delta_2}
|\vec{x}_{13}|^{\Delta_2-\Delta_3-\Delta_1}
|\vec{x}_{23}|^{\Delta_1-\Delta_2-\Delta_3} \,.
\eeq

Finally, a straightforward calculation gives the equal time correlation function of
four renormalised monopole operators with $\sum_{i=1}^4 \alpha_i=0$ in terms of invariant 
cross ratios, 
\begin{align}
\langle \mathcal{O}^R_{\alpha_1}(\vec{x}_1,\tau)\ldots
\mathcal{O}^R_{\alpha_4}(\vec{x}_4,\tau)\rangle
&= \epsilon^{-\widetilde\Delta}
e^{-\frac12 \sum_{i,j=1}^4\alpha_i\alpha_j G_E(\vec{x}_i;\vec{x}_j)} \\
&= 
\prod_{i<j} |\vec x_{ij}|^{\frac{\widetilde\Delta}{3}-\Delta_i-\Delta_j}
X^{\frac{\widetilde\Delta}{3}-\frac{(\alpha_1+\alpha_3)^2}{2^{d+1}\pi^{d/2}\Gamma(d/2) \kappa}}
Y^{\frac{\widetilde\Delta}{3}-\frac{(\alpha_1+\alpha_4)^2}{2^{d+1}\pi^{d/2}\Gamma(d/2) \kappa}} \,,
 \nonumber
 \end{align}
where 
$X=\frac{|\vec x_{12}|\,|\vec x_{34}|}{|\vec x_{13}|\,|\vec x_{24}|}$, 
$Y=\frac{|\vec x_{12}|\,|\vec x_{34}|}{|\vec x_{14}|\,|\vec x_{23}|}$
and $\widetilde\Delta=\sum_{i=1}^4 \Delta_i$.

\subsection{Operators inserted at generic points}
\label{gen2pt}
We now turn our attention to correlation functions in the generalised quantum
Lifshitz model with operators inserted 
at different times as well as different spatial positions. For this we find it 
convenient to adopt a streamlined regularisation procedure along the lines of
the one used by \cite{Ardonne:2003wa} in the original quantum Lifshitz model.
Having established that non-vanishing correlation functions of monopole
operators are independent of the infrared regulator, we can dispense with 
$\mu$ in our formulas, provided we only 
consider correlation functions where the monopole charges
satisfy $\sum_i \alpha_i=0$. The ultraviolet divergences are then 
efficiently handled by introducing a regularised two-point function with the 
equal time coincident point two-point function subtracted off. Starting from 
\eqref{full2ptfcn} we obtain 
\begin{align}
G^R_E(\vec{x}_1,\tau_1&;\vec{x}_2,\tau_2)\equiv
G_E(\vec{x}_1,\tau_1;\vec{x}_2,\tau_2)-
G_E(\vec{x}_1,\tau_1;\vec{x}_2,\tau_2)\Big|_{\{|\vec x_{12}|=\epsilon,\ \tau_{12}=0\}}
\nonumber\\
&=
\frac{1}{2\kappa(2\pi)^{d/2}}\int_0^\infty \frac{dp}{p}\left[
\big(p |\vec{x}_{12}|\big)^{1-\ft{d}{2}} J_{\ft{d}{2}-1}(p |\vec{x}_{12}|)
e^{-\kappa |\tau_{12}|p^d}
-\big(p\epsilon\big)^{1-\ft{d}{2}} J_{\ft{d}{2}-1}(p\epsilon)\right] \nonumber\\
&=\frac{1}{2\kappa(2\pi)^{d/2}}\int_0^\infty dy\,y^{-d/2}
\Big[J_{\ft{d}{2}-1}(y)e^{-\xi y^d}
-\left|\frac{\vec x_{12}}{\epsilon}\right|^{\ft{d}{2}-1}
J_{\ft{d}{2}-1}\big(\frac{\epsilon y}{|\vec x_{12}|}\big)\Big] ,
\label{2pt_integral}
\end{align}
where in the last step we have introduced a scaling variable, 
\beq
\xi \equiv \frac{\kappa |\tau_{12}|}{|\vec x_{12}|^d} \,,
\eeq
which is invariant under the Lifshitz scaling transformation \eqref{xtscaling} with $z=d$.

The equal time correlation functions of renormalised monopole operators are 
easily recovered by setting $\xi=0$ in the
regularised two-point function. In this case the integral in \eqref{2pt_integral} can  
be obtained in closed form and one finds the following exact result for the 
regularised equal time two-point function,
\beq
G^R_E(\vec{x}_1;\vec{x}_2)=-\,\frac{1}{2^d\pi^{d/2}\Gamma\left(\frac{d}{2}\right)\kappa}
\log\Big(\frac{|\vec x_{12}|}{\epsilon}\Big).
\label{equaltime2pt}
\eeq
Wick's theorem then yields the same equal time correlation 
functions of monopole operators as in Section~\ref{equaltimecorrelators}. 

The regularised two-point function at generic points can be expressed as a sum of
two terms: the equal time two-point function \eqref{equaltime2pt} and an integral that  
only depends on the Lifshitz invariant combination $\xi$,
\beq
G^R_E(\vec{x}_1,\tau_1;\vec{x}_2,\tau_2)=
G^R_E(\vec{x}_1;\vec{x}_2)\Big|_{\xi=0}+I_d(\xi)\,,
\label{reg2pt_generic}
\eeq
with
\beq
I_d(\xi)=\frac{1}{2\kappa (2\pi)^{d/2}}
\int_0^\infty dy\,y^{-d/2}J_{\ft{d}{2}-1}(y)\big(e^{-\xi y^d}-1\big)\,.
\eeq
The integral is finite for any finite value of $\xi$ and can easily be evaluated 
numerically for any given number of spatial dimensions. Correlation functions 
of monopole operators inserted at generic points do not have the CFT 
form found for equal time correlators but depend on the Lifshitz invariant
ratio $\xi$ in a non-trivial way. For instance, the correlation function of two
renormalised monopole operators is given by
\beq
\langle \mathcal{O}_{\alpha}^R(\vec{x}_1,\tau_{1})
\mathcal{O}_{-\alpha}^R(\vec{x}_2,\tau_{2})\rangle
=|\vec{x}_{12}|^{-2\Delta}\,e^{\alpha^2 I_d(\xi)}.
\label{eq:generic2pt}
\eeq
In the special case of $d=2$, the integral can be expressed in terms of an incomplete
gamma function
\beq
I_2(\xi)=-\,\frac{1}{8\pi\kappa} \Gamma\big(0,\frac{1}{4\xi}\big) \,,
\eeq
and we reproduce the monopole operator two-point function of \cite{Ardonne:2003wa} 
for the original quantum Lifshitz model, 
\beq
\langle \mathcal{O}_{\alpha}^R(\vec{x}_1,\tau_{1})
\mathcal{O}_{-\alpha}^R(\vec{x}_2,\tau_{2})\rangle
=|\vec{x}_{12}|^{-2\Delta}\,
e^{-\Delta \Gamma\big(0,\frac{|\vec{x}_{12}|^2}{4\kappa|\tau_{12}|}\big)}.
\label{eq:qLifshitz2pt}
\eeq

\subsection{Vacuum autocorrelators}
\label{qlifshitzvacuum}

Next we consider monopole operators located at the same spatial point 
but at different times and evaluate the resulting autocorrelation functions. 
This amounts to taking the limit $|\vec x_{ij}|\rightarrow 0$ for all the 
spatial insertion points before using Wick's theorem to evaluate the 
correlation function of the monopole operators. At first sight, the 
regularised two-point function \eqref{reg2pt_generic} appears singular 
in this limit, due to the logarithmic dependence on $|\vec{x}_{12}|$ in
the equal time two-point function \eqref{equaltime2pt}, 
but this turns out to be cancelled by a corresponding logarithm in the
integral $I_d(\xi)$ at large $\xi$. 

To see this, we differentiate $I_d$ with respect to $\xi$,
\beq
\frac{dI_d}{d\xi} =-\,\frac{1}{2(2\pi)^{d/2}\kappa }
\int_0^\infty dy\,y^{d/2}J_{\ft{d}{2}-1}(y)e^{-\xi y^d}\,,
\eeq
then use the series expansion for the Bessel function,
\begin{equation}
J_m(y) = \left(\frac{y}{2}\right)^{m}\sum\limits_{n=0}^\infty  
\frac{(-1)^n}{\Gamma(n{+}1)\Gamma(m{+}n{+}1)}\left(\frac{y}{2}\right)^{2n},
\end{equation}
and exchange the order of summation and integration to obtain
\beq
\frac{d}{d\xi}I_d =-\,\frac{1}{d\,2^d \pi^{d/2}\kappa}
\sum\limits_{n=0}^\infty  \frac{(-1)^n}{2^{2n}}
\frac{\Gamma(\frac{2n}{d}{+}1)}{\Gamma(n{+}1)\Gamma(n{+}\frac{d}{2})}
\Big(\frac{1}{\xi}\Big)^{\frac{2n}{d}+1}\,.
\eeq
Integrating with respect to $\xi$ and keeping only the leading terms  
at large-$\xi$, we find
\beq
I_d(\xi)=\frac{1}{d\,2^d\pi^{d/2}\Gamma\left(\frac{d}{2}\right)\kappa} 
\Big[-\log \xi+ \widetilde c_d + O(\xi^{-2/d}) \Big] \,,
\eeq
where $\widetilde c_d$ is a $d$-dependent constant of integration. 
Finally, we insert this into \eqref{reg2pt_generic} and see that the logarithm
of $|\vec{x}_{12}|$ is precisely cancelled, leaving a well-defined expression 
for the regularised two-point function of the $\chi$ field inserted at the same spatial 
point at different times,
\beq
G^R_E(\tau_1;\tau_2)=
-\,\frac{1}{d\,2^d\pi^{d/2}\Gamma\left(\frac{d}{2}\right)\kappa} 
\log\big(\frac{\kappa |\tau_{12}|}{e^{\widetilde c_d}\epsilon^d}\big)\,.
\label{zeroT2pt}
\eeq
With the two-point function in hand, the autocorrelation functions of renormalised 
monopole operators, as defined in \eqref{eq:renormoperator}, are easily obtained 
by applying Wick's theorem. For instance, for two monopole operators carrying 
opposite charges inserted at $\vec x$ at times $\tau_1$ and $\tau_2$, 
one finds the following scaling form,
\beq
\langle \mathcal{O}_{\alpha}^R(\vec{x},\tau_1)
\mathcal{O}_{-\alpha}^R(\vec{x},\tau_2)\rangle
=|e^{-\widetilde c_d}\kappa\,\tau_{12}|^{-2\Delta/d}.
\label{eq:vacuum2pt3}
\eeq
The scaling exponent in \eqref{eq:vacuum2pt3} differs from the one 
found in the corresponding equal time correlation function \eqref{eq:vacuum2pt} 
by a factor of $1/d$, reflecting the underlying $z=d$ Lifshitz symmetry.
This result could be anticipated, as the form of
the two-point function of scaling operators is fixed by scale invariance. 
Higher-point functions, on the other hand, are not fixed by scale invariance but
nevertheless higher order autocorrelation functions of monopole operators also
have a characteristic CFT form. For three renormalised monopole operators
with $\alpha_1+\alpha_2+\alpha_3 =0$ we obtain,
\beq
\langle 
 \mathcal{O}_{\alpha_1}^R(\tau_1) 
 \mathcal{O}_{\alpha_2}^R(\tau_2) 
 \mathcal{O}_{\alpha_3}^R(\tau_3)
\rangle
\propto |\tau_{12}|^{(\Delta_3-\Delta_1-\Delta_2)/d}\,
|\tau_{23}|^{(\Delta_1-\Delta_2-\Delta_3)/d}\,
|\tau_{13}|^{(\Delta_2-\Delta_3-\Delta_1)/d},
\eeq
up to an overall constant factor that we have not kept track of. Similarly, the four-point 
autocorrelator with $\sum\alpha_i=0$ 
is given by
\begin{align}
\langle 
 \mathcal{O}_{\alpha_1}^R(\tau_1) 
 \mathcal{O}_{\alpha_2}^R(\tau_2) 
 &\mathcal{O}_{\alpha_3}^R(\tau_3)
 \mathcal{O}_{\alpha_4}^R(\tau_4)
\rangle
\\
&\propto 
\prod_{i<j} |\tau_{ij}|^{\frac{\widetilde\Delta}{3d}-\frac{\Delta_i}{d}
-\frac{\Delta_j}{d}}\,
X^{\frac{\widetilde\Delta}{3d}
-\frac{(\alpha_1+\alpha_3)^2}{d\, 2^{d+1}\pi^{d/2}\Gamma(d/2) \kappa}}\,
Y^{\frac{\Delta}{3d}
-\frac{(\alpha_1+\alpha_3)^2}{d\, 2^{d+1}\pi^{d/2}\Gamma(d/2) \kappa}}\, ,\nonumber
\end{align}
where 
$X=\frac{|\tau_{12}|\,|\tau_{34}|}{|\tau_{13}|\,|\tau_{24}|}$, 
$Y=\frac{|\tau_{12}|\,|\tau_{34}|}{|\tau_{14}|\,|\tau_{23}|}$
and $\widetilde\Delta=\sum_{i=1}^4 \Delta_i$.
The apparent CFT structure clearly generalises to $n$-point autocorrelators for any 
$n$ in the ground state. 

We have seen that both equal time correlation functions and autocorrelation 
functions in the generalised quantum Lifshitz model have the appearance of
CFT correlation functions while this does not hold for correlation functions of
operators inserted at generic points in space and time.
For the autocorrelation functions it is less clear why there should be any 
connection to a CFT as in this case the correlation functions are not simply given 
by path integrals weighted with the vacuum wave functional and they involve time 
evolution with respect to a Lifshitz symmetric rather than a conformally invariant
Hamiltonian. 

The relation to a CFT for autocorrelation functions can be established 
in another way, which reveals that the CFT in question is not the same as 
the one involved for the equal time correlation functions. In fact, the vacuum
autocorrelation functions of the generalised quantum Lifshitz model in any 
number of spatial dimensions are matched by those of a standard free boson
CFT in two Euclidean dimensions. 

Starting from (\ref{eq:twopointfn}) and 
setting $\vec{x}_1=\vec{x}_2$, the two-point function of the elementary  
field is given by
\beq
G_E(\tau_1;\tau_2)=\int \frac{d\omega d^dp}{(2\pi)^{d+1}}
\frac{e^{-i\omega\,\tau_{12}}}{\omega^2+\kappa^2 p^{2d}}.
\label{eq:vacuum2ptauto}
\eeq
Now write the $p$ integral in spherical coordinates, change
variables to $q=\kappa p^{d}$ and extend the range of integration over $q$ to 
be from $-\infty$ to $\infty$, to obtain
\beq
G_E(\tau_1;\tau_2)=
\frac{1}{d\,2^{d}\pi^{d/2} \Gamma\left(\frac{d}{2}\right)\kappa}
\int 
\frac{d\omega dq}{2\pi}\frac{e^{-i\omega\tau_{12}}}{\omega^2+q^2}.
\label{eq:vacuumcft2pt}
\eeq
This is the two-point function at coincident spatial 
points of a free two-dimensional CFT with the action
\beq
S_{CFT}=d\,2^{d-2}\pi^{\frac{d}{2}-1}\Gamma(d/2)\kappa 
\int dy d\tau\Big((\partial_{\tau}\chi)^2+(\partial_y\chi)^2\Big).
\label{eq:CFTaction}
\eeq
The key point in the above reduction is that the integrand in 
\eqref{eq:vacuum2ptauto} only depends on the magnitude of the momentum 
and not on its direction. Thus, it works only for autocorrelators. 

To evaluate the two-point function (\ref{eq:vacuumcft2pt}) in this approach, we would 
introduce infrared and ultraviolet regulators and proceed in parallel with the calculation
presented in Section~\ref{equaltimecorrelators} for equal time correlation functions. 
It is straightforward to check that such a calculation reproduces the autocorrelation 
functions of monopole operators in the generalised quantum Lifshitz model found above.

\subsection{Autocorrelators in a thermal state}
\label{qlifshitzthermal}

We round up our discussion of the generalised quantum Lifshitz 
model at $z=d$ by considering thermal autocorrelation functions. 
In the Matsubara formalism the two-point function of the
$\chi$ field in a thermal state is given in terms of the sum
\beq
G_E(\tau_1,\tau_2)=T\sum_n\int \frac{d^{d}p}{(2\pi)^{d}}
\frac{e^{-i\omega_n\tau_{12}}}{\omega_n^2
+\kappa^2 p^{2d}},\qquad \omega_n=2\pi n T.
\eeq
As before, we make the change of variables in the momentum integral $q=\kappa p^d$, which gives
\beq
G_E(\tau_1,\tau_2)=\frac{2\pi T}{d\, 2^{d}\pi^{d/2} \Gamma(\frac{d}{2})\kappa}
\sum_n\int \frac{dq}{2\pi}\frac{e^{-i\omega_n\tau_{12}}}{\omega_n^2+q^2}.
\label{thermal2pt}
\eeq
This is precisely the thermal two-point function of the free conformal field theory 
with the action (\ref{eq:CFTaction}) \cite{Ghaemi}. The thermal autocorrelation functions of 
monopole operators are thus identical to those of a two-dimensional free boson, 
as they can be obtained from the $\chi$ autocorrelation function using Wick contraction.

We use the same regularization procedure as in Section~\ref{gen2pt} above and restrict 
our attention to correlation functions where the monopole charges satisfy $\sum_i \alpha_i=0$.
A regularised two-point function, with the equal time two-point function subtracted 
off, is given by
\begin{eqnarray}
G^R_E(\tau_1;\tau_2)&\equiv&
G_E(\tau_1;\tau_2)-
G_E(\tau+\tilde\epsilon;\tau) \nonumber\\
&=& - \frac{1}{d\, 2^{d}\pi^{d/2} \Gamma(\frac{d}{2})\kappa}
\log \left[\frac{\vert \sin(\pi T \tau_{12})\vert}{\pi T\tilde\epsilon}\right],
\end{eqnarray} 
where $\tilde\epsilon$ is an ultraviolet cutoff in the Euclidean time direction and we have
performed the integral and sum in \eqref{thermal2pt}. This reduces to the zero 
temperature two-point function in \eqref{zeroT2pt} in the $T\rightarrow 0$ limit
provided the temporal and spatial UV cutoffs are related by 
\beq
\kappa\,\tilde\epsilon = e^{\tilde c_d}\epsilon^d \,.
\eeq
The thermal autocorrelator of two renormalised monopole operators is then given by
\beq
\label{thermaltwo}
\big\langle \mathcal{O}_{\alpha}^R(\vec{x},\tau_1)
\mathcal{O}_{-\alpha}^R(\vec{x},\tau_2)\big\rangle
=\frac{(\pi T)^{2\Delta/d}}{\vert e^{-\tilde c_d}\kappa\,\sin(\pi T\tau_{12})\vert^{2\Delta/d}}.
\eeq
For later reference, we include also the thermal autocorrelator of three renormalised 
monopole operators in the generalized Lifshitz model,
\begin{align}\label{thermalthree}
\langle 
 \mathcal{O}_{\alpha_1}^R(\tau&_1) 
 \mathcal{O}_{\alpha_2}^R(\tau_2) 
 \mathcal{O}_{\alpha_3}^R(\tau_3)
\rangle \\
&\propto\frac{(\pi T)^{(\Delta_1+\Delta_2+\Delta_3)/d}}
{|\sin(\pi T\tau_{12})|^{(\Delta_1+\Delta_2-\Delta_3)/d}\,
|\sin(\pi T\tau_{23})|^{(\Delta_2+\Delta_3-\Delta_1)/d}\,
|\sin(\pi T\tau_{13})|^{(\Delta_3{+}\Delta_1{-}\Delta_2)/d}}.
 \nonumber
 \end{align}

\section{Holographic models with Lifshitz scaling}
\label{holomodel}

Anisotropic scaling of the form \eqref{xtscaling} with $z>1$ is realized in holographic
models through geometries that are asymptotic to the so-called Lifshitz spacetime
\cite{Kachru:2008yh,Koroteev:2007yp}
\beq
ds^2=\ell^2\left(\frac{d\tau^2}{u^{2z}}+\frac{du^2}{u^2}+\frac{d\vec{x}^{\,2}}{u^2}\right),
\label{eq:lifshitzspacetime}
\eeq
Here $\ell$ is a characteristic length scale in the higher dimensional bulk spacetime 
and the coordinates $\tau$, $u$, and $\vec{x}$ are dimensionless. For convenience, 
we adopt units such that $\ell=1$. The Lifshitz metric \eqref{eq:lifshitzspacetime} is 
invariant under 
\beq
\tau\rightarrow \lambda^z\,\tau \,, \qquad \vec{x}\rightarrow \lambda\, \vec{x} \,, \qquad
u \rightarrow \lambda\, u \,,
\label{eq:lifshitzscaling}
\eeq
which incorporates the scaling in \eqref{xtscaling} on the $\tau$ and $\vec{x}$
coordinates. Spacelike infinity is at $u=0$ and the geometry has a null singularity 
at $u\rightarrow \infty$, where tidal forces diverge while all scalar curvature invariants
remain finite. This is a peculiarity of the Lifshitz vacuum spacetime. Finite temperature
states in the dual boundary field theory instead correspond to Lifshitz black holes with
a non-singular horizon at a finite value of $u$.

The Lifshitz spacetime is known to be a solution of the
field equations of several different gravitational models.  The  
Einstein-Maxell-dilaton (EMD) theory \cite{Taylor:2008tg},
\beq
S_\textrm{EMD} = -\int\mathrm{d}^{d+2}x\sqrt{-g}\;\left[
R-2\Lambda-\frac{1}{2}\partial_\mu\phi\partial^\mu\phi
-\frac{1}{4} e^{\alpha\,\phi}F_{\mu\nu}F^{\mu\nu}\right],
\label{eq:emdaction}
\eeq
with negative cosmological constant $\Lambda =-\frac12(d+z)(d+z-1)$
and $\alpha = - \sqrt{2d/(z-1)}$ is a particularly convenient choice. This model
has well known analytic black hole solutions for generic $z\geq 1$ \cite{Tarrio:2011de}, 
which we utilize when we discuss thermal correlation functions in 
Section~\ref{thermalcorrelators} below. 
Our results in the present section, including those higher-order vacuum correlation 
functions in Section~\ref{vacuumautocorrelators}, hold for arbitrary values of $d$ and $z$.
They only rely on the form of the Lifshitz metric \eqref{eq:lifshitzspacetime} and do not 
depend on the choice of gravitational model.

The Lifshitz metric is a solution of the EMD field equations when the dilaton and 
gauge field have the following background values,
\beq
e^\phi=\left(\frac{u_0}{u}\right)^{\sqrt{2d(z-1)}}\,,\qquad
F_{u\tau}=i\sqrt{2(z+d)(z-1)}\,u_0^{-z-1}\left(\frac{u_0}{u}\right)^{d+z+1} ,
\label{vacuumphi}
\eeq
where the factor of $i$ appears due to the Euclidean signature and $u_0$ is an arbitrary reference value of $u$ which arises due to the shift symmetry of the action
\begin{equation}\label{eq:shiftsym}
\phi \rightarrow \phi + c, \qquad A_\mu \rightarrow e^{-c\alpha/2}A_{\mu}.
\end{equation}
As expected when $z= 1$, the dilaton is independent of $u$, the auxilliary gauge 
field vanishes, and the Lifshitz metric reduces to that of anti-de Sitter space.

A $z\geq 1$ planar black brane solution is given by 
\beq
ds^2=f(u)\frac{d\tau^2}{u^{2z}}+\frac{1}{f(u)}\frac{du^2}{u^2}+\frac{d\vec{x}^{\,2}}{u^2},
\label{bhmetric}
\eeq
with 
\beq
f(u)=1-(u/u_{\textit{\scriptsize H}})^{z+d}
\eeq
and the same dilaton and gauge field \eqref{vacuumphi}
as the Lifshitz vacuum. Now there is a horizon at $u=u_{\textit{\scriptsize H}}$ 
and in order for the Euclidean spacetime geometry to be smooth there the time 
$\tau$ must be periodic with a period that 
corresponds to the Hawking temperature 
\beq
T=\frac{d+z}{4\pi u_{\textit{\scriptsize H}}^z} \,.
\label{bhtemp}
\eeq
Due to the underlying scale symmetry of the planar black brane solution, 
all finite temperatures are physically equivalent in this system. Indeed, one
is always free to rescale the horizon radius $u_{\textit{\scriptsize H}}$ to unity by a scale 
transformation of the form \eqref{eq:lifshitzscaling} accompanied by an appropriate shift (\ref{eq:shiftsym}),
upon which the temperature in (\ref{bhtemp}) becomes a pure number.

\subsection{Scalar two-point functions in the geodesic approximation}

Now consider a minimally coupled scalar field with the action
\beq
S=\frac12 \int d^{d+2}x\sqrt{g}\left((\partial\varphi)^2 + m^2 \varphi^2 \right) .
\label{scalaraction}
\eeq
In general, the two-point function of the scaling operator $\mathcal{O}$ in the dual 
boundary field theory, that is dual to the bulk field $\varphi$, is obtained from a
solution of the Klein-Gordon equation,
\beq
(-\nabla^2 + m^2)\varphi =0 \,.
 \label{scalareq}
\eeq
Near the $u\rightarrow 0$ boundary the solutions have the asymptotic form
\beq
\varphi(\tau,\vec{x},u)=\varphi_-(\tau,\vec{x})\, u^{\Delta_-}
+\varphi_+(\tau,\vec{x})\, u^{\Delta_+}+\ldots \,.
 \label{near_boundary}
\eeq
with 
\beq
\Delta_\pm = \frac{d+z}{2}\pm \sqrt{\left(\frac{d+z}{2}\right)^2+m^2} .
 \label{scalingdim}
\eeq
The scaling dimension of $\mathcal{O}$ is $\Delta=\Delta_+$ and for large scalar 
mass we have $\Delta\approx  m\gg 1$. 

The two-point correlation function of operators of high scaling dimension, $\Delta\gg 1$, 
can be expressed in terms of  the length of the shortest bulk geodesic connecting 
the two insertion points $x_1=(\tau_1,\vec{x}_1)$ and $x_2=(\tau_2,\vec{x}_2)$ on the 
boundary~\cite{Hartle:1976tp,Balasubramanian:1999zv},
\beq
\langle\mathcal{O}^R({x}_1)\mathcal{O}^R({x}_2)\rangle\approx \epsilon^{-2 \Delta} 
e^{-\Delta\, L({x}_1;{x}_2)}.
\label{twopointfunction}
\eeq
Here $\epsilon$ is an infrared cutoff in the bulk spacetime, $u>\epsilon$.
The geodesic approximation can be motivated either from a saddle point approximation 
in the particle path integral representation of the bulk Klein-Gordon propagator, or 
from a WKB solution of the bulk Klein-Gordon equation \cite{Festuccia:2005pi}. 

We will determine the geodesic by minimizing the length functional
\beq
L=\int d\lambda\,\frac{1}{2}\Big(e^{-1}g_{\mu\nu}\frac{dx^{\mu}}{d\lambda}
\frac{dx^{\nu}}{d\lambda}+e\Big),
\label{eq:geodesicaction}
\eeq
where $e$ is a vielbein that satisfies the following equation of motion
\beq\label{vielbeineq}
e^2=g_{\mu\nu}\frac{dx^{\mu}}{d\lambda}\frac{dx^{\nu}}{d\lambda}.
\eeq
We can use the coordinate reparametrization symmetry to set $e=1$ so 
that the length functional reduces to
\beq
L=\int d\lambda=\lambda_2-\lambda_1,
\eeq
where $\lambda_1$, $\lambda_2$ are the values of the affine parameter of the 
geodesic at its endpoints on the boundary. 

We first consider vacuum correlation functions of high-dimension operators, 
which are captured by geodesics in the Lifshitz spacetime \eqref{eq:lifshitzspacetime}.  
Thermal correlators obtained from geodesics in Lifshitz black hole backgrounds will be 
considered in Section~\ref{thermalcorrelators}.
We orient the spatial coordinates on the boundary so that the geodesic endpoints to 
lie on the $x$-axis, in which case the geodesic equations become
\beqn
\dot{\tau}&=&E\, u^{2z},\label{eq:E}
\\
\dot{x}&=&p\, u^2,\label{eq:p}
\\
\frac{\dot{u}^2}{u^2}&=&1-p^2u^2-E^2u^{2z}, \label{eq:udot}
\eeqn
where $\dot{f}\equiv df/d\lambda$ and $E$, $p$ are constants. 
Using  \eqref{eq:udot} the affine parameter can be expressed as an integral over
the radial coordinate,
\beq
\lambda = \int \frac{du}{u} \frac{1}{\sqrt{1-p^2u^2-E^2u^{2z}}}+\textrm{constant.}
\label{lambdaintegral}
\eeq
Below, we focus on some special values of $z$, $p$, and $E$ where the integral
can be explicitly evaluated. Numerical results are readily obtained for generic values
of these parameters but we will not pursue that here. 

\subsection{Two-point function at equal time and the two-point autocorrelator}

We are not able to evaluate the integral in (\ref{lambdaintegral}) analytically in full
generality,\footnote{In Appendix \ref{sec:z=2} we consider the special case of $z=2$, 
for which which one can analytically compute the integral in (\ref{lambdaintegral}) 
for both $E$ and $p$ non-zero (for any number of spatial dimensions $d$).} 
but we can proceed by setting either $E=0$ or $p=0$, which corresponds to 
equal time correlation functions or autocorrelation functions, respectively.

We begin by considering a geodesic connecting boundary points that are spatially
separated but at equal times, which amounts to setting $E=0$ in \eqref{lambdaintegral}. 
Adjusting the constant of integration so that $\lambda=0$ 
corresponds to the midpoint of the geodesic, one finds a radial profile 
\beq
u(\lambda) = \frac{1}{\vert p\vert}\,\frac{1}{\cosh{\lambda}}\,.
\label{equaltimeu}
\eeq
With an infrared cutoff the geodesic endpoints are at $u=\epsilon$, which translates
into
\beq
\lambda_{1,2} =  \pm\log\Big(\frac{\epsilon \vert p\vert}{2}\Big)+O(\epsilon^2)
\,.
\label{lambdaendpoints}
\eeq
By integrating \eqref{eq:p} and taking the
$\epsilon\rightarrow 0$ limit, we find that $p=2/|\vec{ x}_{12}|$.
The regularized geodesic length is then given by
\beq
L_R = -2\log{\epsilon}+2\log{\vert\vec{ x}_{12}\vert}+O(\epsilon^2) \,,
\label{equaltimelength}
\eeq
and the equal time vacuum two-point function has the expected scaling form,
\beq
\langle\mathcal{O}^R(\vec{x}_1,\tau)\mathcal{O}^R(\vec{x}_2,\tau)\rangle= 
\vert\vec{ x}_{12}\vert^{-2\Delta} \,,
\label{equaltimetwopoint}
\eeq
which is independent of the value of $d$ and $z$.

The autocorrelator at generic $z$ is obtained by setting $p=0$ in \eqref{lambdaintegral}
and going through the same steps as before. The radial profile of the geodesic is given by
\beq
u = \left(\vert E\vert\cosh{z\lambda}\right)^{-1/z}\,.
\label{uparametrization}
\eeq
and by integrating \eqref{eq:E} we find $E=2/(z|\tau_{12}|)$. The regularized geodesic length is 
\beq
L_R = -2\log\epsilon +\frac2z \log{(z\vert\tau_{12}\vert)}\,.
\label{generalzlength}
\eeq
Inserting this into \eqref{twopointfunction} leads to a scaling form of
the two-point autocorrelator,
\beq
\langle\mathcal{O}^R(\vec{x},\tau_1)\mathcal{O}^R(\vec{x},\tau_2)\rangle
= \left(z\,\vert\tau_{12}\vert\right)^{-2\Delta/z} \,.
\label{generalztwopoint}
\eeq
This expression differs from that of the equal time correlator in precisely the way one 
expects from the Lifshitz scaling relation \eqref{xtscaling}.

\subsection{Three-point vacuum autocorrelators}
\label{vacuumautocorrelators}

In this section we obtain the three-point autocorrelator of large-dimension scalar operators 
in the Lifshitz vacuum, {\it i.e.} a boundary correlation function, with all three operators 
inserted at the same spatial position $\vec{x}=0$ but at different (Euclidean) times 
$\tau_1$, $\tau_2$, $\tau_3$. We consider a bulk theory with a three-point vertex of 
the form
\beq
-\frac{\lambda}{3!}\int d^{d+2}x\sqrt{g}\phi_1(x)\phi_2(x)\phi_3(x),
\eeq
where the fields $\phi_j$ have masses $m_j$. To first order in powers of $\lambda$, 
the three-point function, is given by a tree-level Witten diagram,
\beq
G_3(\tau_1,\tau_2,\tau_3)=-\lambda \int d^dxd\tau du \sqrt{g} 
G^{(1)}_{BB}(0,\tau_1;\vec{x},\tau,u) G^{(2)}_{BB}(0,\tau_2;\vec{x},\tau,u) 
G^{(3)}_{BB}(0,\tau_3;\vec{x},\tau,u) .
\label{eq:3ptwitten}
\eeq
In the limit of large scaling dimensions, $\Delta_j=m_j\gg 1$, the integral can be 
performed in a saddle point approximation,
\beq
G_3(\tau_1,\tau_2,\tau_3)\propto-\lambda \,
e^{-\sum_j \Delta_j \,L_R(0,\tau_j;\vec x,\tau,u)} \,,
\eeq
where $L_R(0,\tau_j;\vec{x},\tau,u)$ is the regularised length of a geodesic 
connecting the boundary point $(0,\tau_j)$ and the bulk point $x^\mu=(\vec{x},\tau,u)$.
The bulk vertex is positioned so as to minimise the sum 
over the lengths of the bulk-to-boundary geodesics (weighted by the 
corresponding scaling dimensions),
\beq
\frac{\partial}{\partial x^{\mu}}\sum_{j=1}^3 \Delta_j  L_R(0,\tau_j;\vec{x},\tau,u)=0.
\label{eq:saddlepteq}
\eeq
It is clear by symmetry that the bulk saddle point will be at $\vec x =0$ and we
only have to vary $\tau$ and $u$ to locate it.

At this point it is convenient to change the radial coordinate to $y=u^z/z$ so
that the Lifshitz metric \eqref{eq:lifshitzspacetime} becomes
\beq
ds^2=\frac{1}{z^2}\frac{1}{y^2}(d\tau^2+dy^2)+\frac{1}{(z y)^{2/z}}d\vec{x}^2,
\eeq
The part of the metric that is relevant for geodesics in the $(\tau,y)$-plane 
is that of AdS$_2$ with a characteristic radius rescaled by a 
factor of $1/z$ compared to that of the Lifshitz spacetime. 
The geodesics are simply arcs of semicircles in the new coordinates and their 
regularised length is given by
\beq
L_R(0,\tau_j,\epsilon^z/z;0,\tau,y)=\frac{1}{z}\left(\log{\left((\tau-\tau_j)^2+y^2\right)}
-\log y + \log z- z\log \epsilon\right) \,.
\label{reglength}
\eeq
The saddle point equations (\ref{eq:saddlepteq}) reduce to
\begin{align}
\sum_{j=1}^3\frac{\Delta_j}{z}\,\frac{\tau-\tau_j}{(\tau-\tau_j)^2+y^2}&=0\,,
\\
\sum_{j=1}^3\frac{\Delta_j}{z}\,\frac{(\tau-\tau_j)^2-y^2}{(\tau-\tau_j)^2+y^2}&=0\,,
\end{align}
and after some algebra one finds the following solution
\begin{align}
y^2&=\frac{\gamma}{4 P^2}\tau_{12}^2\tau_{23}^2\tau_{13}^2\,,\label{eq:solution}
\\ 
\tau&=\frac{X}{2 P}\,,
\end{align}
where
\begin{eqnarray}
\gamma&=&2\Delta_1^2\Delta_2^2+2\Delta_1^2\Delta_3^2+2\Delta_2^2\Delta_3^2-\Delta_1^4-\Delta_2^4-\Delta_3^4\,,\nonumber
\\
P&=&\Delta_1\Delta_2 \tau_{12}^2+\Delta_2\Delta_3 \tau_{23}^2+\Delta_1\Delta_3 \tau_{13}^2
-\Delta_1^2\tau_{12}\tau_{13}-\Delta_2^2\tau_{21}\tau_{23}-\Delta_3^2\tau_{31}\tau_{32}\,,
\\
X&=&\Delta_1^2\tau_{12}\tau_{31}(\tau_2+\tau_3)
+\Delta_2^2\tau_{12}\tau_{23}(\tau_1+\tau_3)
+\Delta_3^2\tau_{23}\tau_{31}(\tau_1+\tau_2)\nonumber \\
&\phantom{=}&+2\Delta_1\Delta_2\tau_{12}^2\tau_3+2\Delta_1\Delta_3\tau_{13}^2\tau_2
+2\Delta_2\Delta_3\tau_{23}^2\tau_1\,.\nonumber
\end{eqnarray}
Non-trivial cancellations occur when the solution is inserted into the expression 
for the regularised length \eqref{reglength}, yielding rather simple time dependence.
For instance, the geodesic connecting the bulk vertex to the boundary insertion at
$\tau_1$ has length
\beq
L_R(0,\tau_1,\epsilon^{z}/z;0,\tau,y)=
\frac{1}{z}\log\Big[\frac{2\Delta_1(\Delta_1-\Delta_2-\Delta_3)}{\sqrt{\gamma}}\Big]
+\frac{1}{z}\log\Big[\frac{|z\tau_{12}||z\tau_{13}|}{|z\tau_{23}|\epsilon^{z}}\Big]\,.
\eeq
The lengths of the geodesics connecting to the boundary at $\tau_2$ and $\tau_3$
are obtained by cyclic permutation. This, in turn, leads to the following CFT-like form for the 
three-point autocorrelator,
\beq
G_3(\tau_1,\tau_2,\tau_3)\propto
\frac{C(\Delta_1,\Delta_2,\Delta_3)}
{|\tau_{12}|^{(\Delta_1+\Delta_2-\Delta_3)/z}
|\tau_{13}|^{(\Delta_1+\Delta_3-\Delta_2)/z}
|\tau_{23}|^{(\Delta_2+\Delta_3-\Delta_1)/z}}\,,
\label{eq:holo3pt}
\eeq
where the analog of the OPE coefficient is given by
\begin{align}\label{coeff}
C(\Delta_1,\Delta_2,\Delta_3)&=-\lambda \,
(\gamma/2)^{\Delta/z}
\Delta_1^{-\Delta_1/z}\Delta_2^{-\Delta_2/z}\Delta_3^{-\Delta_3/z}z^{-(\Delta_{1}+\Delta_{2}+\Delta_{3})/z}
\nonumber
\\
&\times (\Delta_1{-}\Delta_2{-}\Delta_3)^{-\Delta_1/z}
(\Delta_2{-}\Delta_3{-}\Delta_1)^{-\Delta_2/z}
(\Delta_3{-}\Delta_1{-}\Delta_2)^{-\Delta_3/z}.
\end{align}

\section{Holographic thermal autocorrelators}
\label{thermalcorrelators}

In this section we outline the computation of two- and three-point thermal autocorrelators.
The geodesics relevant to the autocorrelators are located on a constant $\vec{x}$ slice, as they provide
an extremum of the geodesic length functional. Thus, the only
part of the metric relevant to the autocorrelator geodesics is
\beq
ds_2^2=g_{\tau\tau}d\tau^2+g_{uu}du^2.
\eeq
On a Lifshitz black brane with $z=d$, this part of the metric has the form
\beq
ds_{2,Lif}^2=\Big(1-\frac{u^{2d}}{u_H^{2d}}\Big)\frac{d\tau^2}{u^{2d}}
+\frac{du^2}{u^2(1-\frac{u^{2d}}{u_H^{2d}})},\label{eq:lifs}
\eeq
corresponding to a thermal state with $T=d/2\pi u_{H}^{d}$. In gravitational duals of 1+1 
dimensional CFTs the spacetime dual to a thermal state is the BTZ black hole. 
The corresponding part of the metric in a BTZ black hole is
\beq
ds_{2,BTZ}^2=\frac{1}{y^2}\Bigg[\Big(1-\frac{y^2}{y_H^2}\Big)d\tau^2
+\frac{dy^2}{1-\frac{y^2}{y_H^2}}\Bigg].\label{eq:btz}
\eeq
The geometric reason for the agreement of the autocorrelators is that there is a time 
independent coordinate transformation relating (\ref{eq:lifs}) and (\ref{eq:btz}) up to a 
constant rescaling of the metric,
\beq
y=\frac{u^d}{d},\quad ds_{2,Lif}^2=\frac{1}{d^{2}}ds_{2,BTZ}^2.\label{eq:metric_relation}
\eeq
The radial positions of the black brane horizons are related by $y_H=u_H^d/d$. Within the geodesic approximation the factor of $1/d^{2}$ in front of the BTZ metric is important in getting right the Lifshitz scaling 
dimension $\Delta\approx m/d$.  This is apparent when we consider the thermal two-point correlation function in the geodesic approximation,
\beq
G_2\propto e^{-m L_{Lif}}= e^{-\frac{m}{d} L_{BTZ}}.
\eeq
The fact that the relevant part of the Lifshitz black hole metric can be transformed into the 
BTZ metric implies agreement between thermal $z=d$ Lifshitz autocorrelators and 
thermal autocorrelators of a 1+1 dimensional CFT for large scaling dimension operators.
In what follows, we check this explicitly for thermal two- and three-point correlation functions.

\subsection{Holographic thermal two-point autocorrelators}

The time translational symmetry of the action (\ref{eq:geodesicaction}) leads to a conserved energy
\beq
E=u^{-2d}\left(1-\frac{u^{2d}}{u_H^{2d}}\right)\dot{\tau}\,,\label{eq:energy}
\eeq
and since we are interested in the autocorrelator we can set $\vec{x}$ to be constant along the geodesic and use $e=1$ in \eqref{vielbeineq}, to solve for $u(\lambda)$,
\beq
\frac{\dot{u}^2}{u^2}=1-\left(u_H^{-2d}+E^2\right)u^{2d}\,.
\eeq
This leads to the integral
\beq
\lambda-\lambda_0=\int\frac{d\tilde{u}}{\tilde{u}\sqrt{1-\tilde{u}^{2d}(1+\tilde{E}^2)}},
\eeq
where $\lambda_0$ is a reference point along the geodesic, which will turn out to be the turning 
point, and we have introduced rescaled variables $\tilde{u}=u/u_H$ and $\tilde{E}=u_H^d E$ 
to simplify notation. This can be integrated to
\beq
\tilde{u}^d(\lambda)=\frac{1}{\sqrt{1+\tilde{E}^2}\cosh\left[d(\lambda-\lambda_0)\right]}.
\eeq
Next we obtain $\tau$ from (\ref{eq:energy}),
\beq
\tilde{\tau}(\lambda)=\int d\lambda\,\frac{\tilde{E}}{(1+\tilde{E}^2)\cosh^2[d(\lambda-\lambda_0)]-1},
\eeq
where we have also introduced a rescaled time variable $\tilde{\tau}=\tau/u_H^d$. 
Performing the integral leads to the identity
\beq
\tan(d\,\tilde{\tau})=\frac{1}{\tilde{E}}\tanh[d(\lambda-\lambda_0)]\,,\label{eq:tan}
\eeq
where we have used the symmetry of the metric under Euclidean time translations to
set $\tau=0$ at the turning point.
Next we require that as $\lambda\rightarrow\lambda_{1,2}$, the radial coordinate 
approaches the cutoff at $u=\epsilon$, which allows us to determine 
$\lambda_1$ and $\lambda_2$ and the regularised length of the geodesic becomes
\beq
L_R=\lambda_2-\lambda_1=\frac{2}{d}\log\left(\frac{2u_H^d}{\epsilon^d
\sqrt{1+\tilde{E}^2}}\right).\label{eq:thermallength}
\eeq
Now the only problem left is to relate $\tilde{E}$ to the time separation between the endpoints 
of the geodesic. This can be obtained from (\ref{eq:tan}) by taking the limit 
$\lambda\rightarrow\lambda_2$ and requiring that $\tau\rightarrow\tau_{2}=\tau_{12}/2$. 
This leads to
\beq
\tilde{E}=\frac{1}{\tan{(\pi T \tau_{12})}}\,,
\eeq
where $T$ is the temperature of the Lifshitz black brane. Plugging this value of $\tilde{E}$ into 
the regularized length in (\ref{eq:thermallength}) and inserting the resulting expression into
\eqref{twopointfunction} finally gives the two-point autocorrelation function,
\beq
\langle\mathcal{O}^R(\vec{x},\tau_1)\mathcal{O}^R(\vec{x},\tau_2)\rangle =
\Bigg(\frac{\pi T}{d\,\sin (\pi T|\tau_{12}|)}\Bigg)^{2\Delta/d},
\label{eq:holographiccorrelator}
\eeq
which reduces to \eqref{generalztwopoint} in the zero temperature limit.

\subsection{Holographic thermal three-point autocorrelators}

To calculate the thermal three-point function it is convenient to use the coordinates
$(y, \tau)$  for the Lifshitz black brane as defined in (\ref{eq:btz}) and (\ref{eq:metric_relation}). 
The regularized length of 
an equal space geodesic connecting the bulk point $(\tau, y, \vec{x} = 0)$ and the boundary point 
$(\tau_i, \vec{x} = 0)$ is then given by the corresponding expression in the BTZ spacetime, 
multiplied by the 
constant factor $1/d$.
\beq
L^{(i)}_R = \frac{1}{d}\log \left[\frac{2d\, y_H^2}{y\epsilon^d}\left(1-\sqrt{f(y)}
\cos\left(\frac{\tau - \tau_i}{y_H}\right)\right)\right],
\eeq
where we denote $f(y) = 1 - \frac{y^2}{y_H^2}$.
The calculation of the three-point function in the geodesic approximation proceeds as 
in the vacuum case, this time extremising the action
\beq
S_R = \sum_{i = 1}^3 \frac{\Delta_i}{d}\log \left[\frac{2d y_H^2}{y\epsilon^d}\left(1-\sqrt{f(y)}\cos\left(\frac{\tau - \tau_i}{y_H}\right)\right)\right].\label{eq:thermal3ptaction}
\eeq
As before, the three-point function is given by the saddle point value, 
$G_3 \approx e^{-S_R}$. 

At this point it is convenient to introduce new coordinates that map the $(\tau, y)$ section 
of BTZ into the Poincare patch of AdS$_2$,
\beq\label{eq:AdS2coords}
\bar{y} = \frac{y}{1 + \sqrt{f(y)}\cos (\frac{\tau}{y_H})},\quad \bar{\tau} 
= \frac{y_H\sqrt{f(y)}\sin (\frac{\tau}{y_H})}{1+\sqrt{f(y)}\cos(\frac{\tau}{y_H})}.
\eeq
In terms of the AdS$_2$ coordinates, the action (\ref{eq:thermal3ptaction}) is given by
\beq
S_R = \sum_{i = 1}^{3}\frac{\Delta_i}{d}\Bigg[\log\Big(\frac{(\bar{\tau} 
- \bar{\tau}_i)^2+\bar{y}^2}{\bar{y}}\Big) -\log\Big( \frac{\bar{\tau}_i^2+y_H^2}{2y_H^2} \Big) + \log d - \log \epsilon^d \Bigg] ,
\eeq
where $\bar{\tau}_i = y_H\tan \frac{\tau_i}{2y_H}$. 
We note that, apart from the second term in the brackets, this expression is the same as the 
action obtained from the regularised geodesic length \eqref{reglength} in the vacuum
(upon setting $z=d$). The extra term arises from the coordinate transformation in 
(\ref{eq:AdS2coords}), but since it does not involve the variables $\bar{\tau}$ and $\bar{y}$ 
the minimisation problem at finite temperature reduces to the corresponding problem in the 
vacuum, which was already solved in section \ref{vacuumautocorrelators}.
Using the result (\ref{eq:holo3pt}), we obtain the thermal three-point function
\beq
G_3(\tau_1,\tau_2,\tau_3) \approx 
\frac{C(\Delta_1,\Delta_2,\Delta_3)\,\prod_{i = 1}^{3}
\Big( \frac{\bar{\tau}_i^2 + y_H^2}{2y_H^2}\Big)^{\Delta_i/d}}
{| \overline{\tau}_{1}-\overline{\tau}_{2}|^{(\Delta_1+\Delta_2-\Delta_3)/d}
| \overline{\tau}_{1}-\overline{\tau}_{3}|^{(\Delta_1+\Delta_3-\Delta_2)/d}
| \overline{\tau}_{2}-\overline{\tau}_{3}|^{(\Delta_2+\Delta_3-\Delta_1)/d}},
\eeq
where $C(\Delta_{1},\Delta_{2},\Delta_{3})$ is given by (\ref{coeff}).
After some algebra this reduces to,
\begin{equation}
\begin{aligned}
G&_3(\tau_1,\tau_2,\tau_3) \\
&\approx
\frac{(\pi T)^{(\Delta_1+\Delta_2+\Delta_3)/d}\,C(\Delta_1, \Delta_2, \Delta_3)}
{|\sin(\pi T\tau_{12})|^{(\Delta_1 + \Delta_2 
-\Delta_3)/d}|\sin(\pi T\tau_{13})|^{(\Delta_1 + \Delta_3 - \Delta_2)/d} 
|\sin (\pi T\tau_{23})|^{(\Delta_2 + \Delta_3 -\Delta_1)/d}} ,
\end{aligned}
\end{equation}
which reduces to \eqref{eq:holo3pt} in the zero temperature limit. 
Up to an overall factor this coincides with the thermal three point function 
(\ref{thermalthree}) of the generalized quantum Lifshitz model, consistent with 
an underlying conformal symmetry.

\section{Excursions outside the time domain}
\label{sec:finite-Delta}

In the previous sections we have seen that, in the limit of large scaling dimensions, the
autocorrelation functions of scalar operators computed in Lifshitz spacetime and Lifshitz 
black brane backgrounds for $z=d$ have the form of autocorrelation functions of a 
$1$+$1$-dimensional conformal field theory. 
In this section we investigate the structure of thermal correlation functions at 
generic values of $\Delta$. We perform the analysis in momentum space and, when possible, 
make contact with the real-space calculations discussed in the previous sections. 

The wave equation for a massive scalar field in the Lifshitz-like black brane background (\ref{bhmetric}) with $z=d$ is 
\begin{equation}\label{eq:d=zwaveeq}
\varphi''(\mfu) -  \frac{1}{1-\mfu}\varphi'(\mfu) + \frac{1}{4\mfu^2(1-\mfu) d^2}
\left(\frac{\mfw^2 \mfu}{(1-\mfu)} - \mfq^2 \mfu^{\frac{1}{d}} 
- \Delta(\Delta-2d)\right)\varphi(\mfu) = 0\,,
\end{equation}
where we have expanded $\varphi$ in Fourier modes as
\begin{equation}
\varphi(\mfu,t,\vec{x}) = e^{-i\omega t+i\vec{p}\cdot\vec{x}}\varphi(\mfu)\,,
\end{equation}
and defined the dimensionless quantities
\beq
\mfu= \frac{u^{2d}}{u_H^{2d}},\qquad\qquad
\mfw = \frac{\omega}{2\pi T/d},\qquad\qquad
\vec{\mfq} = \frac{\vec{p}}{(2\pi T/d)^{1/d}}\,.
\eeq
We wish to determine the analytic structure in frequency space of correlation functions 
of scalar operators dual to the bulk field $\varphi$. We begin by computing the two-point 
correlation function explicitly for $\vec\mfq = 0$ to establish a connection with the 
holographic autocorrelation functions of the previous section. We then determine the 
quasinormal mode frequencies, which correspond to poles of the real-time correlation 
function. We obtain the mode spectrum numerically at generic values of $\mfq$ and also 
analytically in an expansion for small-$\mfq$ as well as in a WKB approximation for large-$\mfq$. 
The details of these calculations are provided in Appendix \ref{app:QNM}, while in the main
text we focus on presenting the results.

\subsection{Correlation functions at $\vec{\mfq} = 0$}

We first consider the situation at zero momentum. As noted in \cite{Sybesma:2015oha}, 
the dependence on the spatial coordinates $\vec{x}$ drops out entirely in this case and 
one can solve for the radial wave-function exactly. The generic solution for $z=d$ can be 
written in terms of hypergeometric functions,
\begin{eqnarray}\label{eq:phiGenSol}
\varphi(\mfu) &=& \varphi_-\, \mfu^{\frac{\Delta_-}{2d}}  
(1-\mfu)^{\beta} \F\left(\ft{\Delta_-}{2d}+\beta,\ft{\Delta_-}{2d}+\beta,\ft{\Delta_-}{d},\mfu \right) \nn\\
&& + \, \varphi_+\, \mfu^{\frac{\Delta_+}{2d}}  (1-\mfu)^{\beta}  \F\left(\ft{\Delta_+}{2d}
+\beta,\ft{\Delta_+}{2d}+\beta,\ft{\Delta_+}{d},\mfu \right),
\end{eqnarray}
where $\Delta_{\pm}$ is defined in (\ref{scalingdim}) and 
\begin{equation}
\beta = \frac{i\mfw}{2d}.
\end{equation}
With the solution to the wave equation in hand, it is straightforward to compute the 
two-point correlation function at $\mfq = 0$ following the prescription in \cite{Son:2002sd}.
The retarded correlation function is given in terms of a flux factor that arises as a boundary 
term when evaluating the (Wick-rotated) action~\eqref{scalaraction} on-shell,
\begin{equation}\label{eq:flux}
\mathcal F(\mfu;\mfw,\mfq) = \frac{1}{2}\varphi_{in}^*(\mfu) \sqrt{g} 
g^{\mfu\mfu} \partial_\mfu \varphi_{in}(\mfu),
\end{equation}
where $\varphi_{in}$ is a solution to the scalar equation of motion which is in-falling at the 
horizon. In terms of $\mathcal F,$ the retarded correlation function is 
\begin{equation}\label{eq:GR}
G^R(\mfw,\mfq) = - 2 \lim\limits_{\mfu\rightarrow\epsilon}\mathcal F(\mfu;\mfw,\mfq),
\end{equation}
where $\epsilon$ is an infrared cut-off on the bulk radial coordinate.
The solution to the scalar wave equation, which is in-falling at the horizon and 
normalized to one at the boundary, is
\begin{equation}\label{eq:PhiSource}
\varphi(\mfu) = \frac{(1-\mfu)^{-\beta}\mfu^{\frac{\Delta_{-}}{2d}}}{(1-\epsilon)^{-\beta}
\epsilon^{\frac{\Delta_{-}}{2d}}} 
\frac{\F(\frac{\Delta_{-}}{2d}-\beta,\frac{\Delta_{-}}{2d}-\beta,1-2\beta,1-\mfu)}
{\F(\frac{\Delta_{-}}{2d}-\beta,\frac{\Delta_{-}}{2d}-\beta,1-2\beta,1-\epsilon)}.
\end{equation}
Using this, one can evaluate (\ref{eq:GR}) to compute the correlation function. 
The main point to make here is that the wave-function in (\ref{eq:PhiSource}) is precisely 
the same as that in \cite{Son:2002sd} for a scalar in the BTZ black hole background at 
zero momentum (with $\Delta_{-}$ replaced by $\Delta_{-}/d$).\footnote{The 
flux-factor $\mathcal F$ is (up to an over-all normalization) the same as that 
computed in \cite{Son:2002sd}. To see this, we note that the coordinate $z_{there}$ 
in \cite{Son:2002sd} naturally generalises to the situation here such that $z_{there} = 1 -\mfu$.} 
Therefore, correlation functions of scalar operators at zero momentum in the $z=d$ Lifshitz 
black brane background are given by zero momentum correlators in a $1$+$1$-dimensional 
conformal field theory (up to a rescaling of all scaling dimensions by a factor $d$).
The reason behind this behaviour is the same as in the previous section. 
In particular, at $\mfq = 0$ the scalar Laplacian is equivalent (after a coordinate 
transformation) to that of a scalar field in a BTZ black hole background, from which 
the relation to the $1$+$1$-dimensional conformal field theory follows.

It is tempting to try to make contact with the autocorrelators discussed in the 
previous sections by performing a Wick-rotation to the Euclidean correlator and then a 
Fourier transform to position-space. Unfortunately, the Fourier transform of the $\mfq = 0 $ 
correlator is not in general equivalent to the autocorrelation function. The autocorrelator is 
instead given by the Fourier transform of the momentum space correlation function, 
computed after first integrating over the spatial momentum,
\begin{equation}
G(\tau_1,\tau_2) = \int \frac{d\omega_E}{2\pi}\int \frac{d^d\vec{p}}{(2\pi)^d} 
e^{-i\omega_E(\tau_1-\tau_2)}  G_E(\omega_E,\vec{p}).
\end{equation}
It is, however, worthwhile to point out that for large $\Delta$ the autocorrelation function 
is dominated by a saddle point which enforces $\vec{p}=0.$ So, at least in this limit, the 
autocorrelator and the $\mfq = 0$ correlation functions are equivalent.  

We have shown that momentum space correlation functions at $\mfq = 0$ in duals to 
the $z=d$ Lifshitz-like black branes (\ref{bhmetric}) are equivalent to those of a $1$+$1$-dimensional 
conformal field theory. In what follows we will investigate what happens when $\mfq \neq 0.$ 
In this case, the scalar wave equation can no longer be solved analytically and we turn to 
computing the quasinormal mode frequency spectrum at $\mfq\ne0,$ which 
corresponds to the poles of the correlation function in frequency space, using a combination of
analytic approximations valid for restricted values of parameters and numerical 
methods for generic parameter values.

\subsection{Quasinormal mode spectrum}

The quasinormal mode spectrum for massive scalar field fluctuations in Lifshitz black brane
backgrounds was computed in \cite{Sybesma:2015oha} for the special case of $\mfq = 0.$ 
An interesting transition was pointed out as one varied the value of $z$ relative to $d$. 
In particular, for $z<d$ both the real and imaginary parts of the quasinormal frequencies 
are non-zero, whereas for $z\ge d$ they become purely imaginary. In \cite{Sybesma:2015oha}, 
the cases $z<d$ and $z\ge d$ were referred to as underdamped and overdamped,
respectively. Furthermore, precisely at $z=d,$ one can determine the spectrum exactly 
to be
\begin{equation}
\mfw_{n,\mfq=0} =  -i(\Delta + 2n d), \qquad n = 0,\,1,\,2,\cdots.
\end{equation}
After the discussion in the previous sub-section it is perhaps not surprising that, up to an 
overall factor of $d,$ this is precisely the quasinormal mode spectrum for $\mfq=0$ in 
a BTZ black hole background. 

\subsubsection{Small-$\mfq$ regime}

We begin by discussing the behaviour of the quasinormal frequencies for small values of 
$\mfq.$ In Table \ref{table:smallk} we present a fit to numerical data for the momentum 
dependence of the lowest quasinormal mode for a massless scalar at several values of 
$z$ and $d.$  

\begin{table}[h]
	\begin{center}
		\begin{tabular}{ | c || c | c | c | c | }
			\hline
			\!\!d\!\!&$z=1$ & $z=2$ & $z=3$  \\
			\hline
			\hline
			\!\!1\!\!&$1.0\mfq-2.0i$ & $-(1.9 + 0.30 \mfq^{2})$$i$ & $-(2.4 + 0.17 \mfq^{2})$$i$  \\
			\!\!2\!\!&$\!1.8\!+\!0.42 \mfq^2\!-\!(2.7\!-\!0.12 \mfq^2)i\!$&$\!0.71 \mfq -(4.0 + 0.17 \mfq^2)i$&$-(3.5 + 0.20 \mfq^2 )i$\\
			\!\!3\!\! &$\!3.1\!+\!0.24 \mfq^2\!-\!(2.7\!-\!0.072 \mfq^2)i\!$&$\!2.1\!+\!0.25 \mfq^2\!-\!(4.9\!+\!0.096 \mfq^2)i\!$&$\!0.58 \mfq\!-\!(6.0\!+\!0.14 \mfq^2)i\!$\\
			\hline
		\end{tabular}\caption{\small Leading momentum dependence (for $\mfq\ll1$) of the lowest quasinormal mode in the spectrum of a massless scalar field for several values of $z$ and $d.$ The $\mfq$-dependence of the entries is obtained by a quadratic polynomial fit to numerical results.}\label{table:smallk}
\end{center}\end{table}

We present the results for $z\neq d$ in order to emphasize the special behaviour that occurs for $z=d.$ In particular, for generic values $z\neq d,$ the behaviour near $\mfq=0$ displays a quadratic dispersion away from the $\mfq = 0$ result and furthermore, the leading dependence on $\mfq$ retains the underdamped (overdamped) behaviour for $z<d$ ($z>d$) of the $\mfq=0$ results.\footnote{Note that the overdamped behaviour for $z>d$ is only robust for a finite region near $\mfq=0.$ Our numerical results (discussed in Appendix~\ref{app:QNM}) indicate that as $\mfq$ increases two quasinormal poles approach each other along the imaginary axis and eventually, at a particular value of $\mfq = \mfq_c$, these two poles merge and obtain a real part for $\mfq > \mfq_c$.} However, for $z=d$ we see that the leading dependence on $\mfq$ is instead linear and real. For the case $z=d=1,$ this is the well-known behaviour for quasinormal frequencies in the BTZ black hole and is exact. For $z=d>1,$ there are two important differences. First, in these cases the linear term is no longer exact and there are further corrections which become important as $\mfq$ increases. Second, although not obvious from this data, the slope of the linear term for $z=d>1$ is a function of both the scaling dimension $\Delta$ of the dual operator and the critical exponent, as opposed to the BTZ result whose linear term has a coefficient precisely equal to one. Nonetheless, the linear approach to $\mfq = 0$ is suggestive and implies that slowly varying spatial perturbations propagate and dissipate in a way similar to a 1+1-dimensional conformal field theory. 

For $z=d$ we can analytically determine the slope by computing the mode function analytically in a  hydrodynamic expansion. Owing to the fact that the $\mfq=0$ mode functions are given exactly by (\ref{eq:phiGenSol}), we can solve the equation (\ref{eq:d=zwaveeq}) perturbatively for small $\mfq$ and determine the quasinormal mode functions and frequencies. We relegate the details of this calculation to appendix \ref{app:QNM}. The leading ($n=0$) quasinormal mode frequency is found to be\footnote{The higher mode frequencies ($n>0$) have similar expressions but the slope depends on $n$ in a nontrivial way. The result for generic $n$ is presented in appendix \ref{app:QNM}.}
\begin{equation}\label{eq:fakesound}
\mfw = -
i \Delta
\pm
\sqrt{\frac{\Gamma\left(\frac{\Delta+1-d}{d}\right)}{\Gamma\left(\frac{1}{d}\right)\Gamma\left(\frac{\Delta}{d}\right)}}\mfq
+
O(\mfq^{2})
\,.
\end{equation}
Note that although in principle one can compute the quasinormal frequencies to generic order in $\mfq,$ for simplicity we contented ourselves with the leading linear-in-$\mfq$ dependence. For massless scalars with $\Delta = 2d,$ this result simplifies to 
\begin{equation}
\mfw_{n=0,\Delta=2d} = - 2 i d\pm \frac{1}{\sqrt{d}}\mfq + \mathcal O(\mfq^2), 
\end{equation}
which matches nicely with the results in Table \ref{table:smallk}.

\begin{figure}[h!]
\begin{center}
\includegraphics[scale=1]{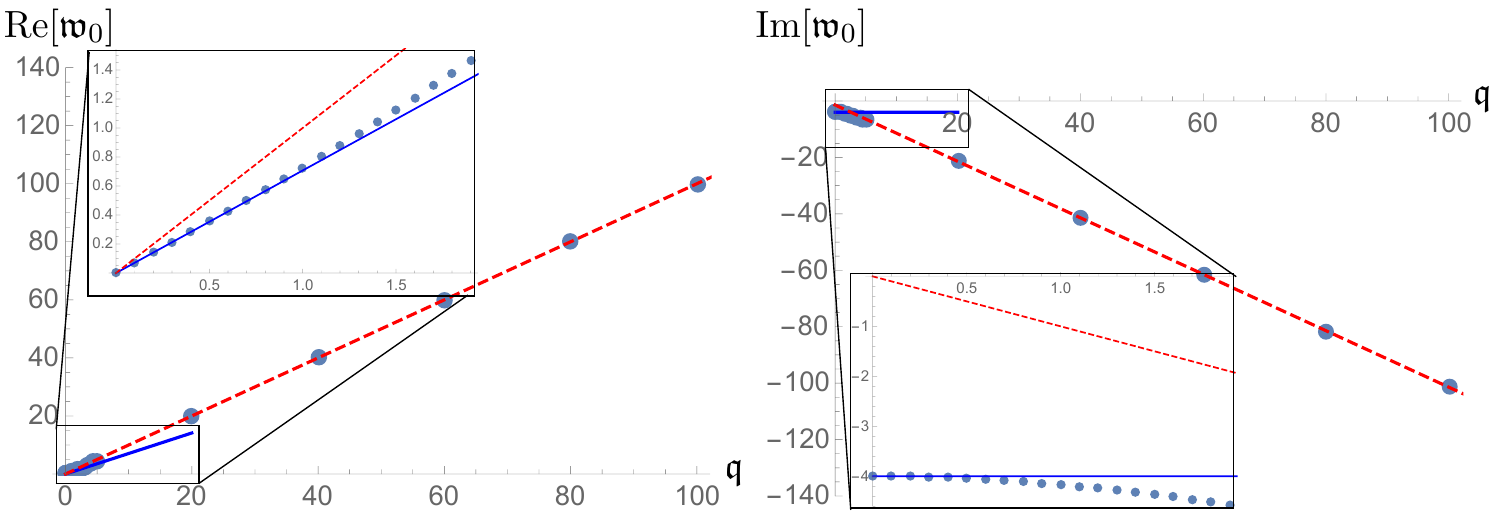}
\caption{\small Comparison between numerical data and analytic results at high and low momentum for the lowest quasinormal mode of a massless scalar in the $z=d=2$ Lifshitz black brane background. The blue dots represent numerical data. The inset illustrates the small-$\mfq$ dependence, where the blue line is the result in (\ref{eq:fakesound}). The red line is the asymptotic large-$\mfq$ result in (\ref{eq:WKBomegadz}).}
	\label{fig:Pevolution}
\end{center}
\end{figure}

Figure~\ref{fig:Pevolution} provides a visualization of the $\mfq$ dependence for the lowest quasinormal mode of a massless scalar. In particular, the insets in the figure highlight the approach to $\mfq=0$ for the real and imaginary parts of the frequency for the special case of $z=d=2.$ The blue dots in the inset refer to numerical data while the blue line is the hydrodynamic result in (\ref{eq:fakesound}). The two are in good agreement at low $\mfq$.

\subsubsection{Large-$\mfq$ regime}

From the data in Figure \ref{fig:Pevolution} it is evident that the quasinormal mode dependence on $\mfq$ becomes linear in $\mfq$ both for small $\mfq$ and large $\mfq.$ These two behaviours are, however, unrelated. As we have seen, the linearity in the small-$\mfq$ regime is specific to the $z=d$ situation and contains information about the approach to equilibrium for slowly varying spatial anisotropy. The linearity at large $\mfq$ is instead generic and occurs for all values of $(z,d).$ 

This can be verified analytically as well. In appendix \ref{app:QNM} we provide a WKB calculation of the linear behaviour for asymptotically large values of $\mfq,$ including the leading off-set from linearity. The result for $z=d$ is given by
\begin{equation}\label{eq:WKBomegadz}
\mfw_n = e^{-i\pi \frac{d-1}{2d}} \sqrt{\frac{d}{d-1}}(d-1)^{\frac{1}{2d}} \mfq - \frac{i d}{\sqrt{2}}(2n+1) + \mathcal O(\mfq^{-1}).
\end{equation}

Finally, it is interesting to note that the large-$\mfq$ results imply that large-momentum excitations in a plasma dual to the Lifshitz-like black brane are always exponentially damped for $z>1.$ This is in stark contrast to the $z=1$ result, where regardless of the dimension, for asymptotically large $\mfq$ the coefficient of the linear-in-$\mfq$ term is exactly $\pm1,$ indicating that high temperature CFT plasmas with holographic duals possess long-lived propagating excitations. In fact, in the $z=1$ case the WKB analysis is slightly more subtle and the overall structure is modified from the above such that the leading correction to the linear term is proportional to $\mfq^{-\frac{d-1}{d+3}}$ and so becomes more suppressed as $\mfq$ increases \cite{Festuccia:2008zx,Fuini:2016qsc}.
\section{Non-equilibrium states}
\label{nonequilibrium}
Up to now we have considered equilibrium states. Both in the $1+1$ CFT case and in the BTZ case it is possible to study the evolution of out of equilibrium states analytically for specific types of quenches. Here we extend those non-equilibrium results to the generalised quantum Lifshitz model and the holographic EMD theory for $z=d$.

\subsection{States with translational and rotational symmetry}\label{qlifshitztranslate}

Autocorrelators in a general translationally and rotationally invariant state of the $z=d$ generalised quantum Lifshitz model are related to autocorrelators in a two-dimensional CFT.
In the Heisenberg picture, the field operator equations of motion are solved by 
(in this section we work in real time)
\beq
\chi(x,t)=\int\frac{d^dp}{(2\pi)^{d}}\frac{1}{\sqrt{2\omega(p)}}\Big(e^{-i\omega(p)t+ip\cdot x}a_p+e^{i\omega(p)t-ip\cdot x}a_p^{\dagger}\Big),
\eeq
where $\omega(p)=\kappa p^d$ and we denote $p^{d}\equiv(p^{2})^{\frac{d}{2}}$. 
The normalization of the operators $a_p$ and $a_p^{\dagger}$ is chosen so that they satisfy
commutation relations $\[a_p,a^{\dagger}_{p'}\]=(2\pi)^{d}\delta^d(p-p')$. 
The Wightman autocorrelator on a general Heisenberg picture state $|\psi\rangle$ is then given by
\begin{align}
&\langle\chi(t_2)\chi(t_1)\rangle=\int\frac{d^dp_1}{(2\pi)^{d}}\int\frac{d^dp_2}{(2\pi)^d}\frac{1}{2\sqrt{\omega(p_1)\omega(p_2)}}\Big(e^{-i\omega(p_1)t_1-i\omega(p_2)t_2}\langle a_{p_2}a_{p_1}\rangle\nonumber
\\
&+e^{i\omega(p_1)t_1-i\omega(p_2)t_2}\langle a_{p_2}a_{p_1}^{\dagger}\rangle+e^{-i\omega(p_1)t_1+i\omega(p_2)t_2}\langle a_{p_2}^{\dagger}a_{p_1}\rangle
+e^{i\omega(p_1)t_1+i\omega(p_2)t_2}\langle a_{p_2}^{\dagger}a_{p_1}^{\dagger}\rangle\Big).
\end{align}
Assuming that the state $|\psi\rangle$ is invariant under spatial translation and rotation leads to the following form of the matrix elements of the creation and annihilation operators
(a proof is given in Appendix \ref{sec:matrixelements})
\begin{align}
\langle a_{p_2}a_{p_1}\rangle= A_{11}(\omega(p_1))(2\pi)^{d}\delta^d(p_1+p_2),\nonumber
\\
\langle a_{p_2}a_{p_1}^{\dagger}\rangle= A_{12}(\omega(p_1))(2\pi)^{d}\delta^d(p_1-p_2),\nonumber
\\
\langle a_{p_2}^{\dagger}a_{p_1}\rangle= A_{21}(\omega(p_1))(2\pi)^{d}\delta^d(p_1-p_2),\label{eq:matrixelements1}
\\
\langle a_{p_2}^{\dagger}a_{p_1}^{\dagger}\rangle= A_{22}(\omega(p_1))(2\pi)^{d}\delta^d(p_1+p_2).\nonumber
\end{align}
The two-point autocorrelator becomes
\begin{align}
&\langle\chi(t_2)\chi(t_1)\rangle=\int\frac{d^dp_1}{(2\pi)^d}\frac{1}{2\omega(p_1)}\Big(e^{-i\omega(p_1)(t_1+t_2)}A_{11}(\omega(p_1))\nonumber
\\
&+e^{i\omega(p_1)(t_1-t_2)}A_{12}(\omega(p_1))+e^{-i\omega(p_1)(t_1-t_2)}A_{21}(\omega(p_1))+e^{i\omega(p_1)(t_1+t_2)}A_{22}(\omega(p_1))\Big).
\end{align}
Using the same change of integration variables as before $q=\kappa p_1^d$ and extending the integration region, we obtain
\begin{align}
&\langle\chi(t_2)\chi(t_1)\rangle=\frac{2\pi}{2^{d}\pi^{d/2}d\kappa \Gamma(\frac{d}{2})}\int\frac{dq}{2\pi}\frac{1}{2\bar{\omega}(q)}\Big(e^{-i\bar{\omega}(q)(t_1+t_2)}A_{11}(\bar{\omega}(q))\nonumber
\\
&+e^{i\omega(p_1)(t_1-t_2)}A_{12}(\bar{\omega}(q))+e^{-i\omega(q)(t_1-t_2)}A_{21}(\bar{\omega}(q))+e^{i\omega(q)(t_1+t_2)}A_{22}(\bar{\omega}(q))\Big),\label{eq:CFTnoneq}
\end{align}
where $\bar{\omega}(q)=|q|$. The autocorrelator (\ref{eq:CFTnoneq}) is identical to the autocorrelator of a 1+1 dimensional CFT with the  action (\ref{eq:CFTaction}), now in Lorentzian time.

The state of a Gaussian CFT is specified by the matrix elements
\begin{align}
\langle b_{q_2}b_{q_1}\rangle=  A_{11}(\bar{\omega}(q_1))2\pi\delta(q_1+q_2),\nonumber
\\
\langle b_{q_2}b_{q_1}^{\dagger}\rangle=  A_{12}(\bar{\omega}(q_1))2\pi\delta(q_1-q_2),\nonumber
\\
\langle b_{q_2}^{\dagger}b_{q_1}\rangle= A_{21}(\bar{\omega}(q_1))2\pi\delta(q_1-q_2),
\\
\langle b_{q_2}^{\dagger}b_{q_1}^{\dagger}\rangle=  A_{22}(\bar{\omega}(q_1))2\pi\delta(q_1+q_2),\nonumber
\end{align}
where the creation and annihilation operators of the CFT satisfy the algebra $\[b_q,b^{\dagger}_{q'}\]=2\pi\delta(q-q')$. 
We have so far demonstrated that the Wightman autocorrelators of the $z=d$ generalised quantum Lifshitz model
are equivalent to the two-dimensional CFT Wightman autocorrelators. Other two-point
correlation functions (such as Feynman or retarded correlators) 
can be obtained from linear combinations 
of Wightman functions and their complex conjugates.
If the wavefunctional of the state $|\psi\rangle$  is Gaussian, 
one can obtain the correlation functions of all the composite 
operators, such as the monopole operators, by using Wick's theorem with 
the $\langle \chi\chi\rangle$  two-point function. Thus, in Gaussian states with spatial 
translational and rotational symmetries, the autocorrelators of the generalised quantum 
Lifshitz model are identical to autocorrelators of a 1+1 dimensional free boson CFT.

\subsection{A mass quench in the generalised quantum Lifshitz model}

As a concrete example of a non-equilibrium state in the generalised quantum Lifshitz model
with $z=d$, we can consider a quench state starting from the ground state of a different 
free Hamiltonian at $t=0$. In this case the matrix elements (\ref{eq:matrixelements1}) are given by
\begin{align}
A_{11}(\omega)=A_{22}=\frac{1}{4}\Big(\frac{\omega}{\omega_0}-\frac{\omega_0}{\omega}\Big),
\\
A_{12}(\omega)=\frac{1}{4}\Big(\frac{\omega}{\omega_0}+\frac{\omega_0}{\omega}+2\Big),
\\
A_{21}(\omega)=\frac{1}{4}\Big(\frac{\omega}{\omega_0}+\frac{\omega_0}{\omega}-2\Big),
\end{align}
where $\omega_0$ is the initial dispersion relation, and $\omega$ the final dispersion relation. 
These matrix elements can be computed from matching the initial and final ground state at $t=0$. 
In the case of a quench starting from the ground state of a mass deformed quantum Lifshitz 
Hamiltonian one has
\beq
\omega_0=\sqrt{\kappa^{2} p^{2d}+m_0^2},\quad \omega=\kappa p^{d}.
\eeq
For simplicity we consider the case of a deep quench with $m_0^{-1}\ll t$ where $t$ 
is any of the following $|t_1-t_2|$, $t_1$ or $t_2$.
This in particular gives the late time behaviour of the correlation function. The full time evolution of 
the correlator can be evaluated numerically. When $m_0$ is large, the leading contribution to the 
two-point function becomes
\beq
\langle\chi(t_2)\chi(t_1)\rangle\approx\frac{2\pi}{2^{d}\pi^{d/2}d\kappa \Gamma(\frac{d}{2})}
\frac{m_0}{8\pi}\int_0^{\infty}\frac{dq}{q^2}\Big(e^{-iq(t_2-t_1)}+e^{iq(t_2-t_1)}
-e^{-iq(t_2+t_1)}-e^{iq(t_2+t_1)}\Big),
\eeq
where we use the integration variable $q=\kappa p^d$. Performing the integrals gives
\beq
\langle\chi(t_2)\chi(t_1)\rangle\approx\frac{2\pi}{2^{d}\pi^{d/2}d\kappa 
\Gamma(\frac{d}{2})}\frac{m_0}{8}\Big(t_1+t_2-|t_2-t_1|\Big).
\eeq
The Feynman propagator is given by
\begin{equation}\begin{aligned}
			G_F(t_2,t_1)
		=&
			\theta(t_2-t_1)\langle\chi(t_2)\chi(t_1)\rangle
			+
			\theta(t_1-t_2)\langle\chi(t_1)\chi(t_2)\rangle
		\\\approx&
		\frac{2\pi}{2^{d}\pi^{d/2}d\kappa \Gamma(\frac{d}{2})}\frac{m_0}{8}
		\Big(t_1+t_2-|t_2-t_1|\Big).
\end{aligned}\end{equation}
The two-point function of monopole operators is obtained by using Wick's theorem
\beq
\langle T e^{i\alpha\chi(t_2)}e^{-i\alpha\chi(t_1)}\rangle\approx e^{-\frac{\pi m_0}{2d}\Delta|t_2-t_1|}.
\eeq
This has the same exponential fall off at large $|t_2-t_1|$ as the thermal 
autocorrelator if one identifies an effective temperature $T_{eff}=m_0/4$.

\subsection{Vaidya collapse spacetime in the holographic model}

A class of non-equilibrium states within holography is provided by 
``quenches" starting from the gapless ground state of the dual field theory. 
This can be achieved by introducing a time dependent source $J(t)$ for some operator 
in the dual field theory. A fast varying source induces a perturbation in some 
(combination of) field(s) close to the boundary of the space, which subsequently falls 
into the bulk and forms a black hole. Generically the time evolution has to be followed
numerically by solving the dynamical Einstein's equations coupled to whatever 
matter is present \cite{Gursoy:2016tgf}. A simple metric that can
be used to model the collapsing configuration is the Vaidya spacetime, which 
corresponds to null and pressureless matter
sourcing Einstein's equations.

An asymptotically Lifshitz version of the Vaidya spacetime was constructed 
in \cite{Keranen:2011xs} and equal time correlators of scalar operators 
obtained in the geodesic approximation. Here we would instead like to consider 
autocorrelators in this time-dependent background.

For $z=d$ the Lifshitz-Vaidya spacetime is given by
\beq
ds^2=-\frac{1}{u^{2d}}\Big(1-m(v) u^{2d}\Big)dv^2-2 \frac{du dv}{u^{d+1}}+\frac{d\textbf{x}^2}{u^2},
\eeq
which can be transformed to the form
\beq
ds^2=\frac{1}{d^{2}}\frac{1}{y^2}\Big[-(1-d^{2}m(v)y^2)dv^2-2 dy dv\Big]
+\frac{1}{(d y)^{2/d}}d\textbf{x}^2,\label{eq:Lifshitz-Vaidya}
\eeq
with the coordinate transformation $y=u^d/d$. The $g_{vv}$ and $g_{yv}$ components 
of the metric (\ref{eq:Lifshitz-Vaidya}) are identical to those
of the BTZ-Vaidya spacetime, and since the large-$\Delta$ limit of the autocorrelation function is insensitive to the 
$d\textbf{x}^2$ part of the metric, they will be identical to the autocorrelator in the 
BTZ-Vaidya background. 
These were computed in \cite{Balasubramanian:2012tu} using the geodesic approximation 
in two different ways with the result
\beq\label{eq:QCorr}
G_F(t_2,t_1)\propto\frac{1}{\Big(t_1\cosh(\pi T t_2)-\frac{1}{\pi T}\sinh(\pi T t_2)\Big)^{2\Delta/d}},
\eeq
where $T$ is the temperature of the final state black hole. Due to the above reasoning, 
this is also the result for the non-equilibrium autocorrelator in the Lifshitz-Vaidya case as well.
In (\ref{eq:QCorr}) we have assumed $t_1<0$ and $t_2>0$. When both times are negative, i.e. before the collapse, the two-point function is instead identical to the vacuum one. Correspondingly, when both times are positive, i.e. after the collapse, the two-point function is identical to the thermal one. 

\section{Summary}
\label{discussion}

In this paper we have studied two types of theories exhibiting a Lifshitz scale invariance:
free field theories and holographic theories. The free field theories generalise 
the well known quantum Lifshitz model to arbitrary number of spatial dimensions. 
These theories can be defined for any value of $z$ but we restrict our attention for the 
most part to models with $z=d$ in order to retain some key properties of the original
quantum Lifshitz model. In particular, precisely for $z=d$, one can define a set of
scaling operators (so-called generalised monopole operators) whose equal time 
correlation functions match those of a $d$-dimensional conformal field theory.
The holographic theories are dual to gravitational models of Einstein-Maxwell-dilaton 
(EMD) type and we consider scalar operators dual to massive scalar fields in the
bulk geometry. By studying the vacuum and thermal correlation functions 
of these two classes of theories we have uncovered several interesting features.

At $z=d$ vacuum autocorrelation functions of scaling operators in the generalized quantum Lifshitz model can be expressed in terms of autocorrelators of a $1{+}1$-dimensional CFT. Likewise, for holographic models a similar relation manifests in the geodesic (or large-$\Delta$) approximation to the autocorrelator. This indicates an enhanced symmetry 
in the time domain that does not follow in an obvious way from the Lifshitz scaling symmetry. 
Furthermore, we find that the relation to a $1{+}1$-dimensional CFT persists when we 
consider autocorrelators in a thermal state. On the gravitational side we expect this 
finite temperature behaviour to be specific to the EMD models and that it will not persist in, 
for instance, Lifshitz models that are dual to bulk theories of Einstein-Proca form.

In the generalised quantum Lifshitz model the relation to autocorrelators in a 
$1{+}1$-dimensional CFT follows from a simple change of variables, $q = \kappa p^d$, 
in the momentum integrals in propagators. On the holographic side, the corresponding
relation can be established by a transformation of the radial coordinate, $y = u^d/d$, 
which maps the $(u, t)$ section of the $z = d$ Lifshitz black brane metric to a 
corresponding $(y, t)$ section of a BTZ black hole metric. 
The relation to a $1{+}1$-dimensional
CFT continues to hold for autocorrelators in out of equilibrium states of the generalised 
Lifshitz model that are invariant under spatial translations and rotations. Similarly,
holographic non-equilibrium states
described by a Lifshitz-Vaidya type collapsing spacetime can be mapped to the 
BTZ-Vaidya spacetime with the same coordinate transformation as in the thermal state. This leads to the equivalence between the large-$\Delta$ limit of autocorrelators in certain non-equilibrium states of holographic 
theories with Lifshitz scaling and $1{+}1$-dimensional CFTs with holographic duals.

We also consider correlation functions of operators with order one scaling dimensions,
{\it i.e.} outside the geodesic approximation. In practice, we look for the poles of retarded 
thermal two-point functions, or in other words the quasinormal mode frequencies 
of the corresponding bulk fields. For zero spatial
momentum, $p=0$, the quasinormal mode spectrum can be found analytically and
agrees with the BTZ quasinormal mode spectrum. This again follows from the same 
coordinate transformation we used when evaluating autocorrelators. Furthermore,
for large scaling dimension operators, the equal space limit of correlation
functions can be argued to correspond to zero spatial momentum.
Thus, we find consistent results with both methods. The quasinormal modes at small
non-zero momenta also share features with the corresponding BTZ quasinormal modes. 
In particular, for $z = d$, we find that the frequency starts out purely real at low enough
momentum and grows linearly with
momentum, as in the BTZ case. 
At high momentum, on the other hand, we find interesting differences compared
to CFT results. In particular, the imaginary part of the quasinormal mode frequency scales 
linearly with $p$ at high momentum. This is in strong contrast to holographic CFTs where 
the imaginary parts approach zero as $p^{-\frac{d - 1}{d + 3}}$ at high momentum. 
This implies that, while CFTs have long lived excitations at high momentum, we find 
that high momentum excitations are very short lived in Lifshitz theories with $z>1$. 
This has interesting implications for thermalisation of far from equilibrium configurations
in these theories, which seems opposite to the pattern of thermalisation found in CFTs, 
where the long lived high momentum excitations are the slowest to equilibrate. 

\section*{Acknowledgements}

We would like to thank A.~Jansen for discussions and clarifications on the Mathematica package ``QNMspectral". 
The research of L.T. was supported in part by Icelandic Research Fund grant 163422-051,
the University of Iceland Research Fund, and the Swedish Research Council under 
contract 621-2014-5838.
The work of W.S. was supported by the Netherlands Organisation for Scientific Research (NWO) 
under the VICI grant 680-47-603. This work is part of the D-ITP consortium, a program of the 
Netherlands Organisation for Scientific Research (NWO) that is funded by the Dutch Ministry 
of Education, Culture and Science (OCW).

\newpage

\begin{appendix}

\section{Ground state wave functional in the generalised quantum Lifshitz model}
\label{appendixA}

The Hamiltonian of the generalised quantum Lifshitz model is given by
\beq
H = \frac{1}{2}\int d^d x\Big( \Pi(x) ^ 2 + \kappa^2 (\nabla ^ z \chi)^2 \Big),
\eeq
where $\Pi(x) = \partial_t\chi$ is the canonical momentum, which in the Schrodinger picture
is replaced by the operator
\beq
\Pi(x) = -i \frac{\delta}{\delta \chi(x)}.
\eeq
The ground state of the theory can be found by solving the time independent Schrodinger
equation
\beq
H\Psi\[\chi\] = E\Psi\[\chi\].
\eeq
At this point if is convenient to introduce the operators
\begin{align}
Q(x) &= \frac{1}{\sqrt{2}}\Big( i\Pi(x) + \kappa (-\Box)^{z/2}\chi(x) \Big),
\\
Q^{\dagger}(x) &= \frac{1}{\sqrt{2}}\Big(- i\Pi(x) + \kappa (-\Box)^{z/2}\chi(x) \Big).
\end{align}
The Hamiltonian can be now expressed as
\beq
H = \int d^dx Q^{\dagger}(x)Q(x) + E_0,
\eeq
where $E_0$ is given by\footnote{Above we have defined $Q^{\dagger}(x)Q(x) = 
\lim_{x'\rightarrow x}Q^{\dagger}(x)Q(x')$.}
\beq
E_0 =\lim_{x'\rightarrow x} \int d^dx \frac{\kappa}{2} (-\Box)^{z/2}\delta^d(x - x') = V_d \int \frac{d^d k}{(2\pi)^d}
\frac{\omega(k)}{2},
\eeq
where $\omega(k) = \kappa (k^2)^{z/2}$. The operator $\tilde{H} = H - E_0$ is
positive definite as for any normalizable state
\beq
\langle \tilde{H}\rangle = \int d^d x \langle \psi|Q^{\dagger}(x)Q(x)|\psi\rangle = \int d^d x ||Q(x)|\psi\rangle||^2 \ge 0.
\eeq
Thus, if we can find a state for which $Q(x)|\psi\rangle = 0$, this state will minimize the energy, i.e. it is the ground
state of the theory. The equation 
\beq
Q(x)\Psi\[\chi\] =\frac{1}{\sqrt{2}} \Big(\frac{\delta}{\delta \chi(x)} + \kappa (-\Box)^{z/2} \chi(x)\Big)\Psi\[\chi\] = 0,
\eeq
can be straightforwardly solved by
\beq
\Psi\[\chi\] = \frac{1}{\sqrt{Z}}e^{-\frac{1}{2}\int d^d x\chi(x)\kappa (-\Box)^{z/2}\chi(x)},
\eeq
where $Z$ is a normalization factor that ensures the wavefunction has unit norm
\beq
Z = \int \[d \chi\] e^{-\kappa\int d^d x\chi(x)(-\Box)^{z/2}\chi(x)}.
\eeq
Thus, the correlation functions of operators at equal time in the ground state are given 
by the expression
\beq
\langle \mathcal{O}(\chi)\rangle =  \frac{1}{Z}\int \[d \chi\] e^{-\kappa\int d^d x\chi(x)(-\Box)^{z/2}\chi(x)}\mathcal{O}(\chi),
\eeq
which can be identified as a correlation functions in a $d$-dimensional Euclidean CFT if $z=d$.

\section{Generic two-point function at $z=2$}\label{sec:z=2}
We want to compute a two-point function in the geodesic approximation with both spatial and temporal dependence. For $z=2$ we can allow for arbitrary values of both $E$ and $p$. In this case \eqref{lambdaintegral} gives
\beq
u^2=\frac{4 e^{2\lambda}}{1+2 e^{2\lambda}p^2+e^{4\lambda}(p^4+4E^2)},
\label{z2parametrization}
\eeq
and is is easy to see that \eqref{equaltimeu} and \eqref{uparametrization} are
recovered by setting $E=0$ or respectively $p=0$ and shifting the affine 
parameter by a constant.
At the endpoints of the regularized geodesic, where $u=\epsilon$, the affine 
parameter takes the values
\beq
\lambda_1=-\log\frac{\epsilon}{2}-\frac{1}{2}\log(p^4+4E^2)+O(\epsilon^2) \,,
\qquad \lambda_2=\log\frac{\epsilon}{2}+O(\epsilon^2)\,,  
\eeq
and the regularized length is given by
\beq
L_R=-2\log\frac{\epsilon}{2}-\frac{1}{2}\log(p^4+4E^2)+O(\epsilon^2)\,.
\label{eq:length1}
\eeq
Inserting this into \eqref{twopointfunction} leads to a simple expression for the
vacuum two-point function of scaling operators in terms of $p$ and $E$,
\beq
\langle\mathcal{O}^R(\vec{x}_1,\tau_1)\mathcal{O}^R(\vec{x}_2,\tau_2)\rangle= 
\left( \Big(\frac{p}{2}\Big)^4+ \Big(\frac{E}{2}\Big)^2\right)^{\Delta/2} .
\label{z2twopoint}
\eeq
Integrating the conservation laws (\ref{eq:E}) and (\ref{eq:p}) gives
\begin{eqnarray}
|\tau_{12}| &=& \frac1E \left(1-\frac{\eta}{\tan\eta}\right)\,, \\
|\vec{x}_{12}| &=&  \frac2p\, \frac{\eta}{\tan\eta}\,.
\end{eqnarray}
where the scaling variable $\eta$ is defined via
\beq
\frac{2E}{p^2}=\tan\eta\,,
\eeq
or equivalently 
\beq
\frac{|\vec{x}_{12}|^2}{|\tau_{12}|}=\frac{2\eta^2}{\tan\eta-\eta}\,.
\label{etarelation}
\eeq
This can in principle be inverted to obtain $\eta$ as a function of
$| \vec{x}_{12}|^2/|\tau_{12}|$.

\begin{figure}[h!]
\begin{center}
\includegraphics[scale=1.2]{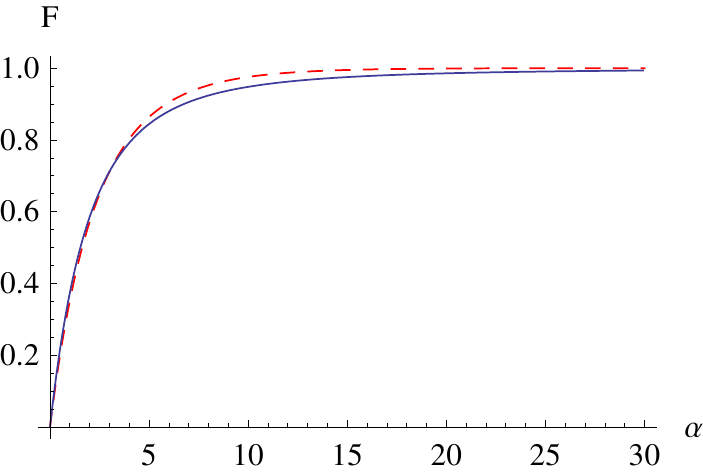}
\caption{\label{fig:comparison} \small The blue (solid) curve shows the scaling function that appears
in the holographic two-point function, while the red (dashed) curve shows the corresponding
scaling function found for two-point functions of monopole operators in the quantum Lifshitz 
model. Both are plotted as functions of the scaling variable $\alpha = |\vec{x}_{12}|^2/|\tau_{12}|$.}
\end{center}
\end{figure}

The vacuum two-point correlation function \eqref{z2twopoint} can be expressed in
terms of the scaling variable in two equivalent ways,
\beq
\langle\mathcal{O}^R(\vec{x}_1,\tau_1)\mathcal{O}^R(\vec{x}_2,\tau_2)\rangle= 
\frac{F(\eta)^\Delta}{\vert\vec{ x}_{12}\vert^{2\Delta}}
=\frac{\tilde{F}(\eta)^\Delta}{\vert2\tau_{12}\vert^\Delta}\,,
\label{eq:holographic}
\eeq
with
\beq
F(\eta)=\frac{\eta^2\cos\eta}{\sin^2\eta}
\qquad\textrm{and}\qquad 
\tilde{F}(\eta)=\frac{1}{\sin\eta}\Big(1- \frac{\eta\cos\eta}{\sin\eta}\Big)\,.
\label{scalingfunctions}
\eeq
We can obtain series expansions for the two-point function in different asymptotic 
limits by using the relation \eqref{etarelation}. 
The limit $|\vec{x}_{12}|^2\gg \vert\tau_{12}\vert$ 
corresponds to $\eta\rightarrow 0$ and 
\beq
\langle\mathcal{O}^R(\vec{x}_1,\tau_1)\mathcal{O}^R(\vec{x}_2,\tau_2)\rangle= 
\frac{1}{\vert\vec{ x}_{12}\vert^{2\Delta}}
\left(1-6\Delta \frac{|\tau_{12}|^{2}}{|\vec{x}_{12}|^4}+\ldots\right),
\eeq
which reduces to the equal time result \eqref{equaltimetwopoint}
when $|\tau_{12}|\rightarrow 0$. On the other hand, $|\tau_{12}|\gg|\vec{ x}_{12}|^{2}$ amounts to 
$\eta\rightarrow \pi/2$ and 
\beq
\langle\mathcal{O}^R(\vec{x}_1,\tau_1)\mathcal{O}^R(\vec{x}_2,\tau_2)\rangle= 
\frac{1}{\vert2\tau_{12}\vert^{\Delta}}
\left(1-\frac\Delta\pi \frac{| \vec{x}_{12}|^2}{\vert\tau_{12}\vert}+\ldots\right),
\eeq
which in turn reduces to the autocorrelator \eqref{generalztwopoint} for $z=2$ when 
$|\vec{ x}|\rightarrow 0$.

The corresponding two-point function can be straightforwardly calculated 
in the original quantum Lifshitz model (see e.g. \cite{Ardonne:2003wa}, 
or our Section \ref{gen2pt}). In the limit
$|\vec{ x}|^2\gg \kappa |\tau_{12}|$ one finds
\beq
\langle\mathcal{O}^R(\vec{x}_1,\tau_1)\mathcal{O}^R(\vec{x}_2,\tau_2)\rangle_\textrm{QL}
= \frac{1}{|\vec {x}_{12}|^{2\Delta}}
\left(1- \frac{4\Delta\kappa |\tau_{12}|}{|\vec{ x}_{12}|^2}
e^{-\frac{ |\vec{x}_{12}|^2}{4\kappa | \tau_{12}|}}+\ldots\right),
\eeq
while the leading behaviour for $ |\vec{x}_{12}|^2\ll \kappa |\tau_{12}|$ is
\beq
\langle\mathcal{O}^R(\vec{x}_1,\tau_1)\mathcal{O}^R(\vec{x}_2,\tau_2)\rangle_\textrm{QL}
=\frac{1}{|2\kappa  \tau_{12}|^{\Delta}}
\left(1-\frac{\Delta}{4}\frac{ |\vec{x}_{12}|^2}{\kappa | \tau_{12}|} +\ldots\right) .
\eeq
The holographic two-point function and the one found in the quantum Lifshitz model agree
in the special cases when either $|\tau_{12}|=0$ or $|\vec{x}_{12}|=0$ and
they both exhibit scaling in $|\vec{ x}_{12}|^2/|\tau_{12}|$ at generic separation, but  
the scaling functions differ in the two theories. In particular, the approach to the equal-time
result is exponentially fast in the quantum Lifshitz model but power law in the holographic model.
The scaling functions are compared in Figure \ref{fig:comparison} where we compare (\ref{scalingfunctions}) and (\ref{eq:qLifshitz2pt}) when plotted against $\alpha\equiv|\vec{x}_{12}|^{2}/|\tau_{12}|$.

\section{Quasinormal mode spectrum}
\label{app:QNM}

In this appendix we provide some details on the computation of the quasinormal mode spectrum. In particular, we elaborate on our numerical procedure as well as obtain analytical results for small-$\mfq$ in (\ref{eq:fakesound}) and large-$\mfq$ in (\ref{eq:WKBomegadz}). 

\subsection{Numerical Algorithm}
To perform the numerical computation we used the Mathematica package ``QNMspectral" \cite{Jansen}. The essence of the algorithm boils down to the following. First we cast the scalar wave equation (\ref{scalareq}) in infalling Eddington-Finkelstein coordinates with a compact radial coordinate, which we discretise on a spectral grid. This grid puts more points near the boundary and horizon than in the ``middle'', which is convenient for numerical purposes. We then express the eigenfunctions in terms of a sum of Chebychev polynomials where we take into account the required fall-off. Note that we can only have as many Chebychev coefficients as points on the spectral grid. This yields a set of equations which can be cast into the form of a generalised eigenvalue problem, 
\begin{equation}
	M\cdot \vec{v}
	=
	\omega\cdot \vec{v}
	\,,
\end{equation}
where $\vec{v}$ is a vector with the coefficients of the Chebychev expansion and $M$ is a matrix with values depending on the details of the quasinormal mode equation (\ref{scalareq}). The eigenvalue problem can readily be solved using Mathematica. In principle, the higher rank the matrix $M$, the higher the precision and number of overtone numbers $n$ of the quasinormal modes $\omega$. The minimum numerical precision of the quasinormal modes computed in this paper is better 
than 1\%.

\subsection{Small-$\mfq$ expansion}

In order to derive (\ref{eq:fakesound}) we begin with the the scalar wave equation in the $z=d$ Lifshitz black brane background (\ref{eq:d=zwaveeq}), which we rewrite here for completeness
\begin{equation}\label{eq:d=zwaveeqAPP}
	\varphi''(\mfu) -  \frac{1}{1-\mfu}\varphi'(\mfu) + \frac{1}{4\mfu^2(1-\mfu) d^2}\left(\frac{\mfw^2 \mfu}{(1-\mfu)} - \mfq^2 \mfu^{\frac{1}{d}} - \Delta(\Delta-2d)\right)\varphi(\mfu) = 0\,.
\end{equation}
For $\mfq=0,$ solution that is regular at infinity is given by (\ref{eq:phiGenSol}) with $\varphi_-=0,$ such that
\begin{equation}\label{eq:phiRegAPP}
\varphi(\mfu) = \varphi_+\, \mfu^{\frac{\Delta_+}{2d}}  (1-\mfu)^{-\frac{i\mfw}{2d}}  \F\left(\ft{\Delta_+}{2d}-\ft{i\mfw}{2d},\ft{\Delta_+}{2d}-\ft{i\mfw}{2d},\ft{\Delta_+}{d},\mfu \right).
\end{equation}
Imposing in-falling boundary conditions at the horizon further restricts the solution such that $\mfw$ is quantized as
\begin{equation}\label{eq:QNMq=0APP}
\mfw_{n,\mfq=0} =  -i(\Delta + 2n d), \qquad n = 0,\,1,\,2,\cdots.
\end{equation}
For generic values of $\mfw$ the hypergeometric function in (\ref{eq:phiRegAPP}) is given by an infinite series in $\mfu,$ however at the quasinormal values of $\mfw$ in (\ref{eq:QNMq=0APP}) the series terminates and one can write the hypergeometric as a finite degree polynomial. In particular, when evaluated on the quasinormal frequency the $\mfq = 0$ scalar solution is (see eq. 15.4.1 of \cite{AbramowitzStegun})
\begin{eqnarray}\label{eq:QNMphi}
\varphi(\mfu) &=& \varphi_+\, \mfu^{\frac{\Delta_+}{2d}}  (1-\mfu)^{-n-\frac{\Delta}{2d}} \F\left(-n,-n,\ft{\Delta_+}{d},\mfu \right) \nn \\
&=& \varphi_+\, \mfu^{\frac{\Delta_+}{2d}}  (1-\mfu)^{-n-\frac{\Delta}{2d}} \sum_{m=0}^{n} \frac{(-n)_m^2}{m!(\ft{\Delta}{d})_m } \mfu^m,
\end{eqnarray}
where the Pochhammer symbol $(\cdot)_m$ is defined by
\begin{equation}
(a)_m \equiv \frac{\Gamma(a+m)}{\Gamma(a)}.
\end{equation}
We will now specialize to the case $n=0$ and will come back to the generic-$n$ situation later.
At $n=0$ the hypergeometric function above is  a constant. In order to find the momentum dependence we then make the following ansatz
\begin{equation}
\varphi(\mfu) = \mfu^{\frac{\Delta}{2d}}(1-\mfu)^{-\frac{\Delta}{2d}-i \frac{\hat\mfw}{2d}}\hat\varphi(\mfu),
\end{equation}
where $\hat\varphi(\mfu)$ parametrizes the deviation from the $\mfq = 0$ solution and we have set $\mfw = - i\Delta + \hat\mfw$,  so that $\hat\mfw$ is the shift of the frequency from the $\mfq=0$ result. Its appearance in the exponent above is required by the in-falling boundary condition at the horizon.

Inserting this into (\ref{eq:d=zwaveeq}) we find the following equation for $\hat \phi(\mfu)$
\begin{equation}
\left(\mfu^{\frac{\Delta}{d}}(1-\mfu)^{1-\frac{\Delta}{d}}\hat\varphi'(\mfu)\right)' + \frac{i\hat\mfw \mfu^{\frac{\Delta}{d}}}{d(1-\mfu)^{\frac{\Delta}{d}}}\hat\varphi'(\mfu) + \frac{\mfu^{\frac{\Delta}{d}-2}}{4d^2(1-\mfu)^{\frac{\Delta}{d}}}(\mfu \hat\mfw^2 - \mfu^{\frac{1}{d}}\mfq^2) \hat\varphi(\mfu)=0.
\end{equation}
Solving this equation perturbatively in $\hat\mfw$ and $\mfq$, and demanding regularity at the boundary $\mfu\rightarrow 0$, one finds that the solution up to $\mathcal O(\hat\mfw^2,\mfq^2)$ is given by
\begin{equation}\begin{aligned}\label{eq:smallqphi}
&\hat{\phi}(\mfu) =
\\
& \phi_0\Bigg[1  - \frac{\Gamma(1-\frac{\Delta}{d})}{4d\Gamma(\frac{1}{d})}\int d\mfu' \mfu'^{-\frac{\Delta}{d}}(1-\mfu')^{\frac{\Delta}{d}-1}\left(\Gamma(1+\ft{\Delta}{d})\Gamma(\ft{1}{d})\hat\mfw^2 - \frac{\Delta}{\Delta+1-d}\Gamma(\ft{\Delta+1}{d})\mfq^2\right) \nn\\
&- \int \frac{d\mfu'}{4d(\Delta-d)}\left(\hat\mfw^2F(1,1,2-\ft{\Delta}{d},1-\mfu') - \mfq^2 \mfu^{\frac{1}{d}-1}F(\ft{1}{d},1,2-\ft{\Delta}{d},1-\mfu')\right) \Bigg] + \mathcal O(\hat\mfw^3,\mfq^3).
\end{aligned}\end{equation}
The important point to note for our purposes is that, when $\frac{\Delta}{d}$ is not an integer,\footnote{In deriving the integrands in (\ref{eq:smallqphi}) we utilized a relation between hypergeometric functions evaluated at $\mfu$ to a sum of two hypergeometric functions evaluated at $1-\mfu$ (see eq. 15.3.6 of \cite{AbramowitzStegun}). For generic values of $\Delta$ one branch simplifies to the elementary functions in the first line. However, when $\Delta$ is a positive integer the connection equation from $\mfu$ to $1-\mfu$ contains terms with $\log(1-\mfu)$ (see eq. 15.3.12 of \cite{AbramowitzStegun}). The vanishing of the logarithmic terms coincides with the condition given in (\ref{eq:smallomegaQNM}).} the integral on the first line gives a solution which is non-analytic at $\mfu=1$ and should be vanishing.  In order for this term to vanish one must set
\begin{equation}\label{eq:smallomegaQNM}
\hat\mfw^2 = \frac{\Gamma(\frac{\Delta+1-d}{d})}{\Gamma(\frac{\Delta}{d})\Gamma(\frac{1}{d})}\mfq^2, 
\end{equation}
which is the result quoted in (\ref{eq:fakesound}). One can perform the same analysis with the generic form of the quasinormal solution in (\ref{eq:QNMphi}) to obtain the general-$n$ result,
\begin{equation}
\hat\mfw^2
=
\frac{\Gamma \left(n+\frac{\Delta }{d}\right)^2}{\Gamma (n+1)^2 \Gamma \left(\frac{\Delta }{d}\right)^2 \Gamma \left(2 n+\frac{\Delta }{d}\right)}	\sum_{i,j=0}^{n}
		\frac{(-n)^{2}_{i}}{\left(\frac{\Delta}{d}\right)_{i}}\frac{1}{i!}
		\frac{(-n)^{2}_{j}}{\left(\frac{\Delta}{d}\right)_{j}}\frac{1}{j!}
		\frac{\Gamma\left[i+j+\frac{\Delta}{d}+\frac{1}{d}-1\right]}{\Gamma\left[i+j+\frac{1}{d}-2n\right]}
		\mfq^2
		\,.
\end{equation}
	
\subsection{Large-$\mfq$ WKB}

At high momentum, we can compute the scalar mode functions in a WKB approximation that is valid for large $\mfq.$ using a method described in \cite{Festuccia:2008zx}. We will carry out the calculation using generic values of $(d,z)$ and later specialize to the case of $d=z.$ 

In order to set notation, let us write the metric for generic $(d,z)$
\begin{equation}
ds^2=\frac{f(u)}{u^{2z}}d\tau^2+\frac{du^2}{u^2f(u)}+\frac{d\textbf{x}^2}{u^2},\quad f(u)=1-\frac{u^{d+z}}{u_H^{d+z}}.
\end{equation}
The scalar wave equation for generic $(d,z)$ is given by
\begin{equation}\label{eomappendix}
\varphi''(\mfu) - \frac{1}{1-\mfu}\varphi'(\mfu) + \frac{1}{\mfu^2(1-\mfu)(d+z)^2}\left(\frac{\mfw^2\mfu^{\frac{2z}{d+z}}}{(1-\mfu)} - \mfq^2 \mfu^{\frac{2}{z+d}} - m^2\right)\varphi(\mfu) = 0,
\end{equation}
where we have again defined dimensionless quantities, which for generic $(d,z)$ are now given by
\begin{eqnarray}
\mfu&=& \frac{u^{d+z}}{u_H^{d+z}},\nn \\
\mfw &=& \left(\frac{4\pi T}{d+z}\right)^{-1}\omega, \nn \\ 
\vec{\mfq} &=& \left(\frac{4\pi T}{d+z}\right)^{-1/z}\vec{p}\,.
\end{eqnarray}
and
\begin{equation}
m^2 = \Delta(\Delta-d-z).
\end{equation}
It is useful to write the wave equation in terms of the so-called ``tortoise" coordinate, which we define to be
\begin{equation}\label{eq:tort}
r_\star = \frac{1}{d+z}\int_0^\mfu \mfu'^{-\frac{d}{d+z}}\frac{d\mfu'}{1-\mfu'}.
\end{equation}
With this convention the boundary is located at $\rs = 0$ and the horizon is approached as $\rs \rightarrow \infty.$ 

In terms of the rescaled wave function $\psi(\mfu) = \mfu^{-\frac{d}{2(d+z)}}\varphi(\mfu)$, the equation of motion (\ref{eomappendix}) can be written in Schr{\"o}dinger form
\begin{equation}
(-\partial_{r_\star}^2 + V(\mfu))\psi = \mfw^2 \psi,
\end{equation}
where the potential is given by
\begin{equation}
V(\mfu) = \frac{1-\mfu}{\mfu^{\frac{2z}{d+z}}}\left[\mfq^2 \mfu^{\frac{2}{d+z}} + \nu^2 - \frac{z^2}{4} + \frac{d^2}{4}\mfu\right],
\end{equation}
with $\nu^2 = m^2 - \frac{(d+z)^2}{4}.$  
To take the large $\mfq$ limit, we rescale the frequency as
\begin{equation}
\mfw = w \mfq,
\end{equation}
in which case the equation of motion can be solved in an expansion in inverse powers of $\mfq.$ In particular, we can write the potential as
\begin{equation}
V(\mfu) = \mfq^2 V_0(\mfu) + V_2(\mfu),
\end{equation}
where 
\begin{equation}\label{eq:WKBV0}
V_0(\mfu) = \mfu^{\frac{2(1-z)}{d+z}}(1-\mfu).
\end{equation}
To determine the leading WKB quasinormal modes, we only need to take into account the leading order potential $V_0(\mfu)$. The additional terms in $V_2(\mfu)$ yield sub-leading contributions that are suppressed by powers of $\mfq^{-1}$. 

In what follows, we will review the analysis of \cite{Festuccia:2008zx}. It is important to note that for $z>1$ the potential in (\ref{eq:WKBV0}) behaves very differently near the boundary compared to the case of $z=1$ studied in \cite{Festuccia:2008zx}. In particular, for $z=1$ (\ref{eq:WKBV0}) approaches a constant at the boundary, whereas for $z>1$ it diverges.\footnote{In fact, as was recently pointed out in \cite{Fuini:2016qsc}, for $z=1$ the WKB analysis does not capture the behaviour of the lowest quasinormal modes for large-$\mfq.$ This is because for $z=1$ the turning point of the potential gets pushed to the near-boundary region, but this does not happen for the $z>1$ case discussed here.} This divergence will naturally lead to bound state solutions. 
The following analysis will therefore echo the large-mass analysis of \cite{Festuccia:2008zx}, rather than the high momentum analysis.

First, consider the situation for real values of $w$, in which case intuition from one-dimensional scattering in quantum mechanics is useful. In particular, since the potential vanishes at the horizon $u=1,$ the function 
\begin{equation}\label{eq:TP}
\kappa(\rs) = w^2 - V_0(\rs)
\end{equation} 
is positive for sufficiently large $\rs$ as one approaches the horizon. The WKB solutions in this region correspond to oscillating waves
\begin{equation}\label{eq:WKBallowed}
\psi_{>,\pm}(\rs) =  \frac{A_{>,\pm}}{\kappa(\rs)^{1/4}} e^{ \pm i \mfq W(\rs) }+ \mathcal O(\mfq^{-1})
\,,
\end{equation}
where
\begin{equation}
W(\rs) \equiv \int_{\rsc}^{\rs} d\rs'\sqrt{w^2-V_0(\mfu(\rs'))}
\end{equation}
and $\rsc$ corresponds to a classical turning point of the potential, such that $\kappa(\rsc) = 0.$ 

As one moves further from the horizon eventually $\rs = \rsc$, after which one enters the classically forbidden region with $\kappa(\rs) < 0$. The WKB solutions in this region correspond to rising and falling exponentials 
\begin{equation}\label{eq:WKBforb}
\psi_{<,\pm}(\rs) = \frac{A_{<,\pm}}{(-\kappa(\rs))^{1/4}} e^{ \pm \mfq \mathcal{Z}(\rs) }+ \mathcal O(\mfq^{-1})\,,
\end{equation}
where
\begin{equation}
\mathcal{Z}(\rs) \equiv \int_{\rsc}^{\rs} d\rs'\sqrt{V_0(\rs')-w^2}.
\end{equation}
Writing the integral out explicitly in terms of the coordinate $\mfu,$ 
\begin{equation}
\mathcal{Z}(u) = \frac{1}{d+z}\int_{\mfu_c}^\mfu d\mfu'\frac{1}{1-\mfu'}\mfu'^{-\frac{d}{d+z}}\sqrt{V_0(\mfu') - w^2},
\end{equation}
where $\mfu<\mfu_c,$ we see that $\mathcal Z(\mfu)$ is negative and decreasing as $\mfu$ moves away from the turning point and so the solution  $\psi_{<,+}(\rs)$ corresponds to the decaying mode under the potential barrier. As one further approaches the boundary at $\rs=0$ the WKB solutions behave as
\begin{equation}
\psi_{<,\pm}(\mfu) \simeq A_{<,\pm} e^{\pm \mfq \mathcal Z(\rsc,0)} \mfu^{\frac{(z-1)}{2(d+z)}} e^{\pm \mfq \mfu^{\frac{1}{d+z}} }.
\end{equation}

In order to construct quasinormal solutions we will build the solution up from the boundary. Near the boundary, the WKB expansion breaks down and one must solve for the wave-function in a near-boundary expansion. At large-$\mfq$, the $w^2$ term is suppressed by powers of $\mfq^{-1}$ and the near-boundary solution is independent of $w$ to lowest order. The regular solution near the boundary is given by
\begin{equation}
\psi_{n.b.}(u) = B \,\mfu^{\frac{z}{2(d+z)}} I_{\nu}(\mfq \mfu^{\frac{1}{d+z}}),
\end{equation}
which scales as $\phi \sim u^{\nu + (d+z)/2}$ for small $u$ as is appropriate for a normalizable mode. To match this to the WKB solutions in (\ref{eq:WKBforb}) we expand for large $\mfu,$  
\begin{equation}
\psi_{n.b.}(u) \simeq B \,\frac{\mfu^{\frac{z-1}{2(d+z)}}}{\sqrt{2\pi \mfq}} e^{\mfq \mfu^{\frac{1}{d+z}}}.
\end{equation}
We therefore see that the normalizable solution at the boundary can be matched onto the decaying WKB solution $\psi_{<,+}(\rs),$ with $A_{<,-}= 0$ and 
\begin{equation}
B =  \sqrt{2\pi\mfq} e^{\mfq \mathcal Z(\rsc,0)} A_{<,+}.
\end{equation}
Next, matching the solution $\psi_{<,+}(\rs)$ across the turning point onto the oscillating solutions in (\ref{eq:WKBallowed}) one finds the solution in the allowed region ($\rs > \rsc$) to have $A_{>,-} = e^{i\pi/2} A_{>,+},$ such that
\begin{equation}
\psi(\rs) = \frac{2A_{>,+}}{\kappa(\rs)^{1/4}} \cos \left(\mfq W(\rs)- \frac{\pi}{4}\right), \qquad \rs > \rsc.
\end{equation}
For large $\rs$ this solution behaves as a linear combination of 
$e^{+ i \mfq \rs}$ and $ e^{- i \mfq \rs}.$ The first behaviour corresponds to an in-falling wave and the latter to an out-going wave. Solutions with real $w$ are therefore not quasinormal. 

To find quasinormal frequencies one must analytically continue by allowing both $w$ and $\rs$ to take complex values. This allows for additional turning points where $\kappa(\rs)$ vanishes. In particular, there will be special complex values of $w$ where one of these new turning points, which we will refer to as $\rst,$ will merge with the analytic continuation of the physical turning point $\rsc.$ As explained in \cite{Festuccia:2008zx}, when this occurs certain sub-dominant contributions to the out-going mode will become of the same order as the leading term. Near the point where the turning points merge, the ratio of the two contributions to the out-going mode is given by
\begin{equation}
e^{2\mfq \mathcal Z(\rsc,\rst)},
\end{equation}
where 
\begin{equation}\label{eq:BSint}
\mathcal Z(\rsc,\rst) = \int_{\rsc}^{\rst} d\rs' \sqrt{V_0(\rs')  - w^2}.
\end{equation}
This means that when 
\begin{equation}
e^{2\mfq \mathcal Z(\rsc,\rst)} = -1,
\end{equation}
the two contributions will exactly cancel and leads to the quantization condition \cite{Festuccia:2008zx}
\begin{equation}\label{eq:WKBquant}
2\mfq \mathcal Z(\rsc,\rst) = i \pi (2n+1), \qquad n = 0,\, 1,\,2,\, \cdots.
\end{equation}

The two turning points merge when the following two conditions are satisfied,
\begin{eqnarray}
V_0(\mfu_b) &=& w_b^2, \nn\\
V_0'(\mfu_b) &=& 0,
\end{eqnarray}
which ensure that $\kappa(\rs)$ has a second order zero at $\rsb \equiv \rs(u_b).$ We can approximate the potential around the point where the two turning points merge as a quadratic polynomial
\begin{equation}
V_0(\rs) \simeq V_0(\rsb) + \frac{1}{2}V_0''(\rsb)(\rs-\rsb)^2.
\end{equation}
Furthermore, assuming $w = w_b + x,$ where $x$ is small, we see that the turning points $\rst$ and $\rsc$ are located at $\rsb \pm a,$ where
\begin{eqnarray}
a = \sqrt{\frac{4w_b x}{V_0''(\rsb)}}.
\end{eqnarray}
In this approximation we can perform the integral
\begin{eqnarray}
\mathcal Z(\rsc,\rst) &=& \int_{\rsc}^{\rst} d\rs'\sqrt{V_0(\rs')-w^2} \nn \\
&\simeq& \int_{-a}^a dy \sqrt{\ft{1}{2}V_0''(\rsb)y^2 - 2 x w_b}\nn\\
&\simeq&  \pm i a^2 \sqrt{\frac{V_0''(\rsb)}{2}}\int_{-1}^1 dy\sqrt{1-y^2} \nn\\
&\simeq& \pm i \frac{a^2\pi}{2}\sqrt{\frac{V_0''(\rsb)}{2}}\nn\\
&\simeq & \pm \frac{i\pi x}{\delta},
\end{eqnarray}
where 
\begin{equation}\label{eq:offset}
\delta = \sqrt{\frac{V_0''(\rs)}{2V_0(\rs)}}\Big|_{\rs=\rsb}.
\end{equation}
The quantization condition (\ref{eq:WKBquant}) then leads to the expression
\begin{eqnarray}
w_n &=& w_b + x \nn\\
&=&w_b + \frac{\delta}{2\mfq}(2n+1).
\end{eqnarray}
The quasinormal frequencies are then given by
\begin{equation}
\mfw = \mfq w_b + \frac{\delta}{2}(2n+1), \qquad n = 0,\,1,\,2,\,\cdots.
\end{equation}

Evaluating this for our potential in (\ref{eq:WKBV0}), we find the turning point conditions to be
\begin{eqnarray}
\mfu_b &=& -\frac{2(z-1)}{d+2-z},\nn\\
w_b &=& e^{-i\pi\frac{(z-1)}{d+z}}\sqrt{\frac{d+z}{d+2-z}}\left(\frac{2(z-1)}{d+2-z}\right)^{-\frac{(z-1)}{d+z}}.
\end{eqnarray}
Evaluating the offset $\delta$ in (\ref{eq:offset}) for this case we find
\begin{equation}
\delta = e^{-i\pi \frac{z}{d+z}} \frac{d+z}{\sqrt{2}}\left(\frac{2(z-1)}{d+2-z}\right)^{\frac{d-z}{2(d+z)}}.
\end{equation}
Note that we only trust these results for $z$ in the range $1<z<d+2.$ The case of $z\ge d+2$ needs to be treated with more care as the point $u_b$, where the turning points merge, goes behind the horizon, {\it i.e.} $\mfu_b>1$. 

\begin{figure}[h!]
\begin{center}
\includegraphics[scale=.9]{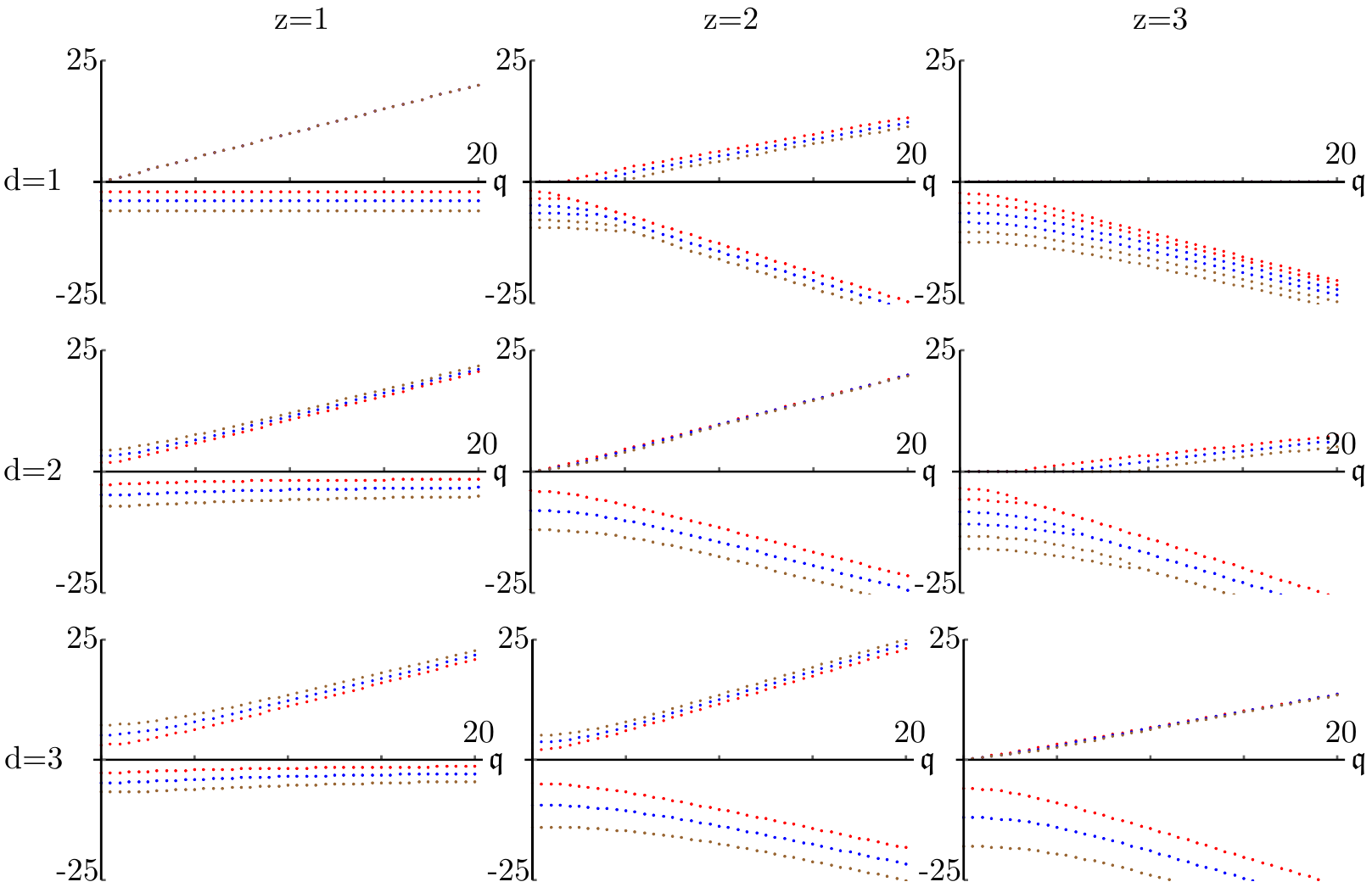}
\caption{\small
                	Real and imaginary parts of the first three overtones ($n=0,1,2$) in the EMD model for a massless field with $\Delta=d+z$. The dots above the x-axis represent the real part and the ones below the axis represent the imaginary part. Notice the different behaviour at small $\mfq$ when $d>z$, $d=z$ or $d<z$. In the $d=1$ and $z=3$ plot, the real parts remain zero, as expected from the large-$\mfq$ WKB result.}\label{numerick}
\end{center}
\end{figure}

For the case of $d=z,$ this gives the following result for the quasinormal frequencies
\begin{equation}
\mfw_n = e^{-i\pi \frac{d-1}{2d}} \sqrt{\frac{d}{d-1}}(d-1)^{\frac{1}{2d}} \mfq - \frac{i d}{\sqrt{2}}(2n+1) + \mathcal O(\mfq^{-1}).
\end{equation}
Figure \ref{numerick} provides an overview of our numerical results for several values
of $d$, $z$ and $\mfq$.
\vskip 18pt

\section{Matrix elements of creation and annihilation operators}\label{sec:matrixelements}

In this Appendix we demonstrate how the momentum dependence of the creation 
and annihilation operator matrix elements (\ref{eq:matrixelements1}) is
fixed by rotational and translational invariance of the state $|\psi\rangle$. 
The vacuum $|0\rangle$ is a state invariant under rotations and translations
\beq
e^{ix\cdot P}|0\rangle=|0\rangle,\quad U(R)|0\rangle=|0\rangle,
\eeq
where $P$ is the momentum operator and $U(R)$ is an operator generating the rotation $R$. 
The one particle state $|p\rangle=a^{\dagger}_p|0\rangle$
transforms as
\beq
e^{ix\cdot P}|p\rangle=e^{ix\cdot p}|p\rangle,\quad U(R)|p\rangle=|Rp\rangle,
\eeq
where $(Rp)^{i}=R^{i}_jp^j$ is the rotated momentum. 
This leads to the following transformation properties of the creation operators
\beq
e^{ix\cdot P}a_p^{\dagger}e^{-ix\cdot P}=a_p^{\dagger} e^{ix\cdot p},\quad 
U(R)a_{p}^{\dagger}U^{\dagger}(R)=a_{Rp}^{\dagger}.
\eeq
The transformation properties of the annihilation operators follow from Hermitian conjugation.
Next we consider one of the matrix elements in (\ref{eq:matrixelements1})
\beq
f(p_1,p_2)=\langle\psi|a_{p_1}a_{p_2}|\psi\rangle.\label{eq:f}
\eeq
We assume that the state $\psi$ is invariant under translations and rotations, 
$e^{ix\cdot P}|\psi\rangle=|\psi\rangle$ and
$U(R)|\psi\rangle=|\psi\rangle$. Inserting three unit operators 
$1=U^{\dagger}(R)U(R)$ in (\ref{eq:f}) gives
\beq
f(p_1,p_2)=f(Rp_1,Rp_2),
\eeq
which means that $f(p_1,p_2)$ is a function of the rotationally invariant 
combinations $p_1^2$, $p_2^2$ and $p_1\cdot p_2$.
By inserting three unit operators $1=e^{-ix\cdot P}e^{ix\cdot P}$ in (\ref{eq:f}) gives
\beq
f(p_1,p_2)=e^{i x\cdot(p_1+p_2)}f(p_1,p_2).
\eeq
Since $f(p_1,p_2)$ should be independent of the arbitrary transformation parameter $x$, $f$ must be of the form
\beq
f(p_1,p_2)= g(p_1)\delta^d(p_1+p_2).
\eeq
Now because of rotational invariance $g(p_1)$ is only a function of $p_1^2$. Finally $p_1^2$ dependence can be traded
to dependence on $\omega(k_1)=\kappa p_1^d\equiv \kappa \left(p_1^{2}\right)^{\frac{d}{2}}$. The same analysis can be repeated for all of the matrix elements in (\ref{eq:matrixelements1}).

\end{appendix}

\end{document}